\documentclass[a4paper,fleqn]{cas-sc}

\usepackage{lineno}
\newcommand*\patchAmsMathEnvironmentForLineno[1]{
  \expandafter\let\csname old#1\expandafter\endcsname\csname #1\endcsname
  \expandafter\let\csname oldend#1\expandafter\endcsname\csname end#1\endcsname
  \renewenvironment{#1}
     {\linenomath\csname old#1\endcsname}
     {\csname oldend#1\endcsname\endlinenomath}}
\newcommand*\patchBothAmsMathEnvironmentsForLineno[1]{
  \patchAmsMathEnvironmentForLineno{#1}
  \patchAmsMathEnvironmentForLineno{#1*}}
\AtBeginDocument{
\patchBothAmsMathEnvironmentsForLineno{equation}
\patchBothAmsMathEnvironmentsForLineno{align}
\patchBothAmsMathEnvironmentsForLineno{flalign}
\patchBothAmsMathEnvironmentsForLineno{alignat}
\patchBothAmsMathEnvironmentsForLineno{gather}
\patchBothAmsMathEnvironmentsForLineno{multline}
}

\usepackage[authoryear,longnamesfirst]{natbib}
\usepackage{physics}
\usepackage{adjustbox}
\usepackage{amsmath}
\usepackage{amssymb}
\usepackage{amsfonts}
\usepackage{color}

\def\tsc#1{\csdef{#1}{\textsc{\lowercase{#1}}\xspace}}
\tsc{WGM}
\tsc{QE}

\begin{document}

\let\WriteBookmarks\relax
\def\floatpagepagefraction{1}
\def\textpagefraction{.001}

\shorttitle{Dynamical Fate of Tidal Fragments}    

\shortauthors{Torii et al.}  

\title[mode = title]{Ring formation around giant planets by tidal disruption of a single passing large Kuiper belt object II: The dynamical fate of tidal fragments}

\affiliation[1]{organization={Department of Earth and Planetary Sciences, Institute of Science Tokyo},
            addressline={Ookayama}, 
            city={Meguro-ku, Tokyo},
            postcode={152-8551}, 
            country={Japan}}

\affiliation[2]{organization={Earth-Life Science Institute, Institute of Science Tokyo},
            addressline={Ookayama}, 
            city={Meguro-ku, Tokyo},
            postcode={152-8550}, 
            country={Japan}}
            
\affiliation[3]{organization={Universit\'{e} Paris Cit\'{e}, Institut de Physique du Globe de Paris, CNRS}, addressline={F-75005}, city={Paris}, country={France}}

\affiliation[4]{organization={Graduate School of Artificial Intelligence and Science, Rikkyo University}, city={Tokyo}, postcode={171-8501}, country={Japan}}

\affiliation[5]{organization={SpaceData Inc.}}

\author[1,2]{Naoya Torii}[orcid=0009-0003-5452-7473]
\ead{torii@elsi.jp}

\author[1,2]{Shigeru Ida}
\ead{ida@elsi.jp}

\author[2,3,4,5]{Ryuki Hyodo}[orcid=0000-0003-4590-0988]
\ead{hyodo@elsi.jp}


\begin{abstract}
Planetary rings are ubiquitous structure in our Solar System, but their formation mechanisms remain under debate. 
One of the proposed scenarios is the tidal disruption of a nearby passing body that enters within a planet’s Roche limit, producing fragments that are gravitationally captured and finally form the rings. 
In this study, we investigate the detailed dynamical path and fate of such tidally captured fragments using direct $N$-body simulations including collisional fragmentation with analytical arguments.
Focusing on Saturn as a representative case, we explore how the inclination ($i_{\rm TD}$) and pericenter distance ($q_{\rm TD}$) of the orbit of the passing body control the subsequent orbital evolution, collisional grinding, and the survival of fragments mass. 
Our simulations show that initially highly eccentric and inclined fragments experience differential precession driven by the planet’s $J_2$ potential, followed by destructive high-velocity collisions that damp their eccentricities and inclinations. 
The timing and pathway of this evolution strongly depend on $i_{\rm TD}$, modifying the dynamical picture proposed in the previous work. 
For low to moderate $i_{\rm TD}$, a narrow, circular and equatorial rings finally form whose orbital radius is well predicted by an analytically derived equivalent circular radius based on the conservation of the vertical component of angular momentum. 
In contrast, for high $i_{\rm TD}$, collisional damping causes a substantial fraction of the material to fall onto the planet, preventing the formation of a massive ring. 
We compile our results of $N$-body simulations with the analytical predictions on $(q_{\rm TD}, i_{\rm TD})$ parameter space and specify the parameter region where sufficient mass to form Saturn’s present rings and inner satellites survives. 
Our results provide a unified dynamical framework linking tidal disruption events, ring formation, and the initial conditions for ring-satellite system evolution, which are also readily applied to the other giant and terrestrial planets.
\end{abstract}


\maketitle

\section{Introduction}\label{sec:Introduction}
The question of how planetary rings formed is a long-standing problem in the planetary science.
In our Solar System, there are wide variety of ring systems around celestial bodies \citep[for review, see][]{Tiscareno:2018, Hyodo:2025b}.
For example, Saturn has massive rings with an interesting dynamical structures including gaps, propellers and density waves \citep[e.g.,][]{Porco:2005, Tiscareno:2007, Hedman:2010, Tiscareno:2019, Torii:2024, Torii:2025}. 
Cassini observations revealed that its main rings are composed of $\gtrsim 95\%$ water ice by mass \citep{Zhang:2017a, Zhang:2017b}, shedding light on Saturn's rings age \citep{Kempf:2023, Durisen:2023, Estrada:2023, Crida:2019, Hyodo:2025a}.
On the other hand, Uranus has several narrow rings radially confined by shepherding moons \citep[e.g.,][]{Elliot:1977, Nicholson:1978, French:2023}, while Neptune has dusty rings with distinct azimuthally confined arcs \citep[e.g., ][]{Hubbard:1986, Nicholson:1990}.
In contrast to Saturn, the rings of Uranus and Neptune are spectrally dark and likely to be higher rock composition \citep{Tiscareno:2013}.
In addition, recent occultation observations revealed that not only giant planets but also some trans-Neptunian objects (TNO) and Centaur objects also have ring systems with narrow ringlets \citep{Braga-Ribas:2014, Ortiz2017, Morgado:2023, Pereira2025}, which is considered to be dynamically confined by the mean-motion resonances with the satellites and/or spin-orbit resonances with the central object. 
This diversity of ring systems indicates that the rings are common feature in our Solar System and these formation scenarios may also be diverse in nature.

So far, several ideas \citep[e.g.,][]{Canup:2010, Hyodo:2017, Dubinski:2019, Wisdom:2022, Teodoro:2023} have been proposed for the formation of ring systems around giant planets \citep[for review of ring origin scenario, see][]{Crida:2025}.
While most of these scenarios focused on the formation of Saturn's rings \citep{Canup:2010, Dubinski:2019, Wisdom:2022, Teodoro:2023}, the tidal disruption of a nearby passing Kuiper belt object \citep{Dones:1991, Hyodo:2017} is applicable not only to Saturn but also to the other giant planets (Jupiter, Uranus and Neptune).
In the tidal disruption scenario, we assume that a small body on a hyperbolic orbit around a planet enters inside its Roche limit.
The passing body is strongly elongated by planet tidal force.
As a result, it is tidally disrupted if the pericenter distance is small enough.
Some fraction of its fragments are gravitationally captured on bound orbit around the planet with an extremely high eccentricity, and they finally form the rings \citep[see Fig. 1 in][]{Hyodo:2017}.
In this paper, we focus on this scenario.

\cite{Charnoz:2009} evaluated the likelihood of this scenario by quantifying the cometary flux around giant planets during the Late Heavy Bombardment (LHB) based on the original simulation of the so-called ``Nice model'' for the origin of the LHB \citep{Tsiganis:2005, Gomes:2005}.
They revealed that the flux was large enough to supply a ring mass in all giant planets in the Solar System.
\cite{Hyodo:2017} also estimated the expected number of tidal disruption events experienced by the giant planets following the procedure in \cite{Charnoz:2009}.
They predicted that Jupiter, Saturn, Uranus and Neptune have experienced a close encounter with Pluto-sized bodies within their Roche limit at least $\sim 8, \ 4, \ 2, \ 2$ times, respectively, assuming the existence of 1000 to 4000 Pluto-sized bodies in the primordial Kuiper Belt \citep{Nesvorny:2016}.
These calculations show that the ring formation by the tidal disruption is viable for the all giant planets in our Solar System at least from the probabilistic point of view.
We highlight that it is important to understand the dynamical fate of captured fragments after the tidal disruption because it is an ubiquitous event in our Solar System
\footnote{
We note that, considering such a high cometary flux around \textit{all} giant planets in the LHB phase, the fact that \textit{only} Saturn has massive and broad rings in the present Solar System should remain open question.
Another mechanism to create the diversity of the ring system around the giant planets may be required \citep{Charnoz:2009}. 
}.

\cite{Dones:1991} derived the analytic formula of the captured mass around the planet during the tidal disruption event assuming uniform energy distribution of the fragments.
However, they neglected the effect of spin and self-gravity of the disrupted body, which have significant effect on the capture efficiency.
In order to refine Dones's formula, \cite{Hyodo:2017} performed SPH simulations of the tidal disruption of a large Kuiper belt object (KBO) with mass of $10^{21}$ and $10^{23}$ kg around Saturn and Uranus, considering the effect of its spin and differentiation.
They found that about $0.1-10\%$ mass of the passing body can be gravitationally captured, which is massive enough to form the current rings and inner regular satellites around Saturn and Uranus \citep[see][]{Crida:2012}.
In addition, they found that the compositional difference between rings of Saturn and Uranus could be naturally explained.
They considered the tidal disruption of a differentiated body with rocky core and icy mantle and showed that a small fraction of silicate originated from the rocky core can be captured as well as icy material.
They found that the silicate fraction of captured material is higher in the case of Uranus than that of Saturn.
This is because the passing objects can enter deep into the Uranus' Roche limit due to the higher density of Uranus and such objects experience more significant tidal disruption, resulting in much higher silicate fraction in the captured fragments originated from its rocky core than the case of Saturn.

\cite{Hyodo:2017} (Paper I) ended their SPH simulations after achieving a steady state in the system and discussed the following long-term dynamical evolution of captured fragments with analytical arguments and non-collisional $N$-body simulation.
Their SPH simulations showed that the eccentricity $e$ of captured fragments is extremely high ($e\gtrsim 0.9$) just after the tidal disruption.
With non-collisional $N$-body simulation, they demonstrated the precession of the argument of pericenter $\omega$ and the longitude of ascending node $\Omega$ of the fragments due to the $J_2$ term of the planet's gravitational potential.
They argued that, from an analytical comparison between the precession and collision timescales, destructive high-velocity collisions ($\sim$ a few km/s) between fragments occur after $\omega$ and $\Omega$ are randomized by their differential precession.
Such energetic collisions generate smaller fragments and damp their eccentricity and inclination, thus circular and thin equatorial rings finally form.
However, no previous studies have conducted a direct collisional $N$-body simulation of the orbital evolution of captured fragments including the collisional fragmentation.

In this work, we examine the detailed orbital evolution of captured fragments by tidal disruption using $N$-body simulations considering self-gravity and collisional fragmentation.
As mentioned above, \cite{Hyodo:2017} demonstrated the orbital precession of fragments without the effects of collisions and discussed its effects separately with a simple analytical arguments.
However, whether both processes --- collisions and precession --- can be treated separately is not obvious, because it is possible for collisions to occur during the orbital precession.
If destructive collisions occur and many fragments are generated during the precession, the collision timescale changes and significant collisional cascade could start.
Thus, the simple timescale comparison used in \cite{Hyodo:2017} seems to be oversimplified.
In order to understand the detail dynamical path after the tidal disruption and the structure of resulting ring, we, for the first time, perform $N$-body simulations including collisional fragmentation.

The structure of this paper is as follows.
We describe our $N$-body simulation methods including the collision outcome model and initial condition in Section \ref{sec:Method},
After that, in Section \ref{sec:Analytical Argument}, we summarize the analytical arguments on the dynamical path after the tidal disruption and equivalent circular orbital radius. 
Section \ref{sec:Simulation Results} consists of our main results of our $N$-body simulations demonstrating the analytical prediction described in Section \ref{sec:Analytical Argument}.
Finally, we compile our $N$-body simulations and discuss the classification of the dynamical path and fate of the captured fragments on a parameter space.
We discuss the collision velocity and several implications on the ring and satellite formation around planets in Section \ref{sec:Discussion} and summarize our findings in Section \ref{sec:Summary}.

\section{Simulation Methods} \label{sec:Method}
We use the open-sourse $N$-body simulation code GPLUM based on Framework for Developing Particle Simulator (FDPS) \citep{Iwasawa:2020, Ishigaki:2021}, which was developed for studying the planetary accretion.
This code uses the particle-particle particle-tree (P$^3$T) scheme for the time integration \citep{Oshino:2011}.
In the original GPLUM code, simple collision outcome model is implemented by default, thus we can treat the collisional evolution including fragmentation following the tidal disruption.

\subsection{Collision outcome model}
In this code, the collision is detected when the physical particle radius overlaps.
If the collision is detected, its outcome is determined based on the outcome model described below.
If the fragments are produced, they are ejected right after the collision event.

Our collision outcome model is the same as that of \cite{Chambers:2013}.
We briefly summarize the model as follow.
Our model includes three regimes of the outcome: (i) fragmentation, (ii) perfect merging, and (iii) hit-and-run.
The outcome for each collision are determined depending on the impact energy per unit mass $Q$ and collision angle $\theta_{\rm col}$ \citep{Leinhardt:2009, Leinhardt:2012}.
If we consider a collision at velocity $v_{\rm col}$ between a target of mass $M_{\rm t}$ and a projectile of mass $M_{\rm p}$ ($M_{\rm t}\geq M_{\rm p}$), the impact energy per unit mass is given by
\begin{equation}
    Q=\frac{\mu v_{\rm col}^2}{2(M_{\rm t}+M_{\rm p})},
\end{equation}
where $\mu$ is the reduced mass defined as
\begin{equation}
    \mu=\frac{M_{\rm t}M_{\rm p}}{M_{\rm t}+M_{\rm p}}.
\end{equation}
If we consider the gravity regime, the catastrophic disruption threshold value of $Q$ for oblique impact, $Q^*$, is given by \citep{Leinhardt:2012}
\begin{equation}
    Q^*=\qty(\frac{\mu}{\mu_{\alpha}})^{3/2}\qty[\frac{(1+\gamma)^2}{4\gamma}]Q_0^*,
\end{equation}
where $\mu_{\alpha}$ is the reduced mass involved in the collision given by
\begin{equation}
    \mu_{\alpha}=\frac{\alpha M_{\rm p}M_{\rm t}}{\alpha M_{\rm p} + M_{\rm t}},
\end{equation}
$\gamma$ is the mass ratio between the target and projectile,
\begin{equation}
    \gamma=\frac{M_{\rm p}}{M_{\rm t}},
\end{equation}
and $\alpha$ is the mass fraction of the projectile involved in the collision \citep[see Eq. (11) in][]{Leinhardt:2012}.
The expression of $Q_0^*$ is given by
\begin{equation}
    Q_0^*=0.8C^*\rho_1GR_{\rm C1}^2,
\end{equation}
where $C^*=1.8$, $\rho_1$ is the density of the impactor (same as that of the target), $G$ is the gravity constant, and $R_{\rm C1}$ is given by
\begin{equation}
    M_{\rm t}+M_{\rm p}=\frac{4\pi\rho_1}{3}R_{\rm C1}^3.
\end{equation}
Given $Q$ and $Q^*$, then we can obtain the largest remnant mass, $M_l$, from the collision \citep{Leinhardt:2012}:
\begin{align}
    M_l&=(M_{\rm t} + M_{\rm p})\qty(1-\frac{Q}{2Q^*}) \ \ \ (Q<1.8Q^*)\notag\\
    &=\frac{(M_{\rm t} + M_{\rm p})}{10}\qty(\frac{Q}{1.8Q^*})^{-3/2} \ \ \ (Q>1.8Q^*). \label{eq:LargestRemnant}
\end{align}

The condition of hit-and-run collision is given by
\begin{equation}
    \sin\theta_{\rm col}>\frac{R_{\rm t}}{R_{\rm t} + R_{\rm p}}, \label{eq:hit-and-run}
\end{equation}
where $R_{\rm t}$ and $R_{\rm p}$ are the radius of target and projectile, respectively \citep[e.g.,][]{Asphaug:2010}.
The condition of merging \citep{Genda:2012} is given by
\begin{equation}
    \frac{v}{v_{\rm esc}} < 2.43\Gamma(1-\sin\theta_{\rm col})^{5/2} -0.0408\Gamma + 1.86(1-\sin\theta_{\rm col})^{5/2}+1.08, \label{eq:merge}
\end{equation}
where $v_{\rm esc}$ is the mutual escape velocity of the target and projectile, and
\begin{equation}
    \Gamma = \qty(\frac{1-\gamma}{1+\gamma})^2.
\end{equation}

\begin{figure}
\centering
\includegraphics[width=0.5\linewidth]{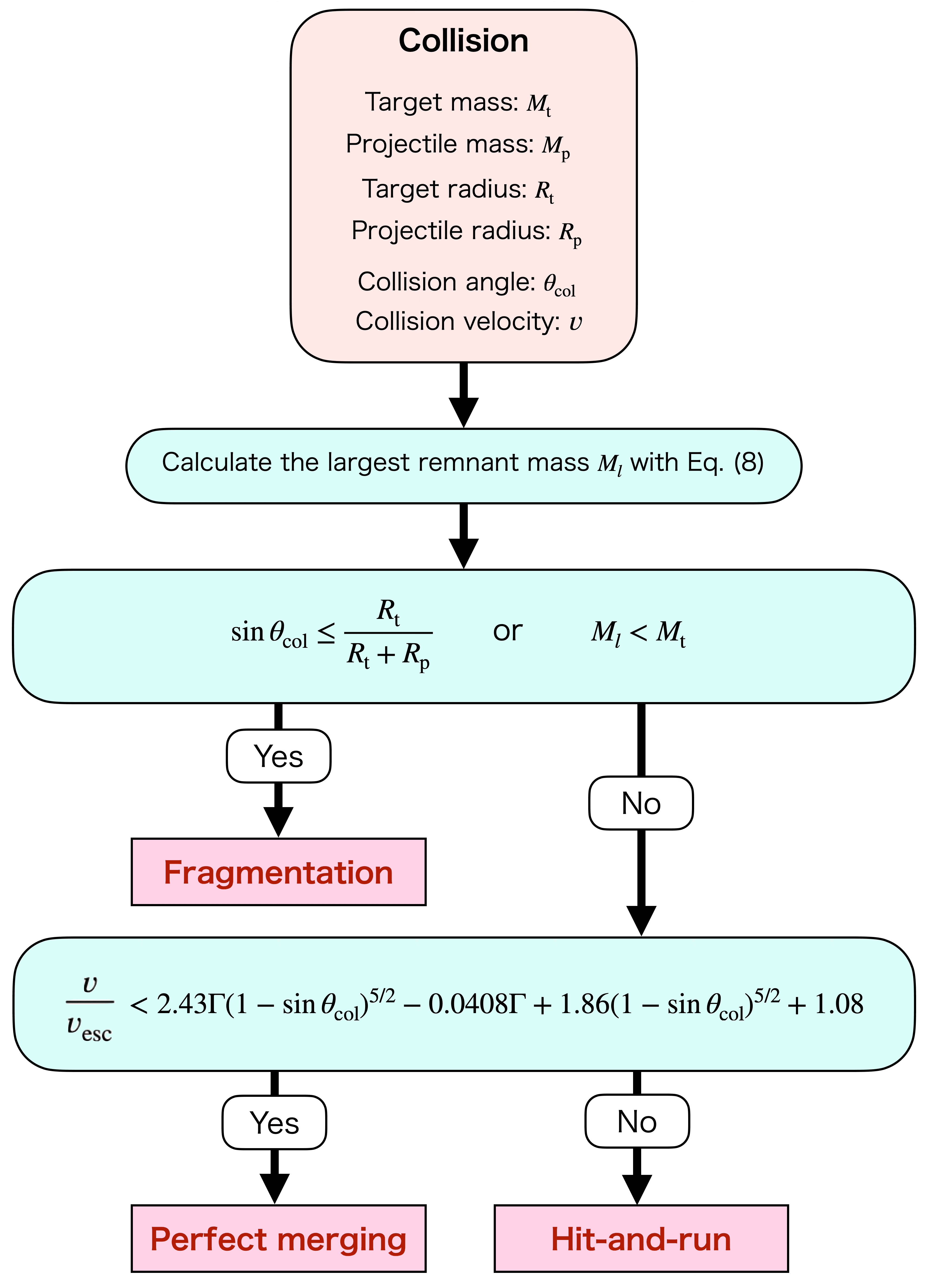}
\caption{The flow chart of our collision outcome model.}
\label{fig:CollisionModel}
\end{figure}

When a collision is detected, the largest remnant mass ($M_l$) is calculated with Eq. \eqref{eq:LargestRemnant} at first.
Then, with the calculated largest remnant mass, the collision outcome is classified based on the above conditions.
We depict the procedure of our collision outcome model in the flow chart of Fig. \ref{fig:CollisionModel}.

\begin{enumerate}
    \item
    If the hit-and-run condition Eq. \eqref{eq:hit-and-run} is not satisfied or the largest remnant mass calculated with Eq. \eqref{eq:LargestRemnant} is smaller than the target mass ($M_l < M_{\rm t}$), the collision is in the fragmentation regime
    \footnote{
    Following \cite{Chambers:2013}, the erosive hit-and-run collision (i.e., the hit-and-run condition Eq. \eqref{eq:hit-and-run} is satisfied and the largest remnant mass is smaller than the target mass $M_l < M_{\rm t}$) is treated as fragmentation regime in our collision model for the ease of implementation.
    }.
    In this regime, the remaining mass $M_{\rm rem}=M_{\rm t} + M_{\rm p} - M_l$ is divided into equal-mass fragments. 
    We set the minimum mass of particles, $M_{\rm min}$, and the maximum number of fragments for one collision, $N_{\rm frag, max}$, to avoid the number of particles in the simulation infinitely increasing.
    Thus, the number of fragments per one collision is given by
    \begin{equation}
        N_{\rm frag} = {\rm min}\qty(\qty[\frac{M_{\rm rem}}{M_{\rm min}}], N_{\rm frag, max}),
    \end{equation}
    where $[x]$ is the floor function which takes the greatest integer less than or equal to $x$.
    Then, the mass of each fragment is given by $M_{\rm rem}/N_{\rm frag}$.
    If the remaining mass $M_{\rm rem}$ is smaller than $M_{\rm min}$ (that is, $N_{\rm frag}=0$), the collision is treated as perfect merging regardless of collision velocity. 
    The largest remnant remains at the center of mass, while the fragments are ejected from launching points near the largest remnant at the relative velocities of 1.05 times the escape velocity of the remnant, because we aim to model the fragments which avoid reaccumulating to the largest remnant, which follows the procedure of \cite{Chambers:2013}. More sophisticated treatment of the ejected fragments should be investigated, but it is out of scope of our study.
    \item
    If the collision is not in the fragmentation regime but the condition Eq. \eqref{eq:merge} is satisfied, the collision is in the merging regime.
    In this regime, the collision is assumed to be perfect merging.
    \item
    If the collision is not in the above two regimes (i.e., when the condition Eq. \eqref{eq:hit-and-run} and $M_l \geq M_{\rm t}$ is satisfied and the condition Eq. \eqref{eq:merge} is not satisfied), the collision is in the (ideal) hit-and-run regime.
    In this regime, the largest remnant and fragments generated from the projectile with mass calculated in a similar way to the above are considered, while the target is not affected by hit-and-run collision \citep[see][for detail procedure]{Chambers:2013}.
\end{enumerate}

The parameters in our collisional model are $N_{\rm frag, max}$ and $M_{\rm min}$.
In the fiducial runs, we set $N_{\rm frag, max}=10$ and $M_{\rm min}=3\times10^{-10}M_0$ corresponding to the mass of $\sim 1.7\times10^{17} \ {\rm kg}$ and the size of $\sim 35 \ {\rm km}$, where $M_0$ is the mass of the central planet, and collisions between the smallest particles with mass $M_{\rm min}$ is assumed to be a perfect merging.
We note that these parameters are introduced to avoid our simulations being too expensive due to producing an excessive number of particles.
However, in reality, destructive collisions could generate fragments more than $N_{\rm frag, max}$ and smaller than $M_{\rm min}$.
We have confirmed by test simulations with $N_{\rm frag, max}=10, 30, 50$ and $M_{\rm min}=3\times10^{-10}M_0, 1\times10^{-10}M_0$ that changing the adopted value of these parameters does not affect the collisional evolution path at least within the simulated timescale and the parameter range, and our conclusion in this paper does not depend on these parameters.
Also, when the eccentricity and inclination of fragments damp and the system shrinks close to the Roche limit, this treatment of collision outcome may be no longer valid and the collision should be treated as an inelastic collision.
Thus, our simulation does not treat the ring viscosity originated from inelastic collisions \citep{Araki:1986} and self-gravity wakes \citep{Daisaka2001} properly.
This means that it cannot handle the long-term viscous evolution of the formed circular ring, while the initial evolution where high-speed collisions frequently occur between particles with high eccentricity can be appropriately treated.
Nevertheless, in the following sections, we will show that our main conclusion are not modified by the details of collision outcome model because the timescale we consider here is much shorter than the viscous timescale and the viscous effect is negligible at least in our simulated timescale.

\subsection{Simulation settings}\label{sec:settings}
In our $N$-body simulation, we consider the central planet as Saturn, but our results could be easily applied to the other giant and terrestrial planets as long as the treatment in the gravity regime of the collision outcome model is valid (see Section \ref{sec:Inclination}).
Saturn is located at the origin and its axisymmetric gravitational potential including up to $J_2$ term is given by \citep{Murray:1999}
\begin{equation}
    \Phi(r, \theta, \phi)=-\frac{GM_0}{r} + \frac{GM_0J_2R_0^2}{2r^3}(3\cos^2\theta - 1) \label{eq:gravitational potential},
\end{equation}
where ($r, \theta, \phi$) is the spherical coordinate and $R_0$ is the radius of Saturn. 
In our simulation, the particles which fall onto the surface of Saturn or reaches at the distance of $r=1000 R_0\sim 2r_{\rm H, Saturn}$ from the planet are removed, where $r_{\rm H, Saturn}$ is the Hill radius of Saturn.

The mass considered here is $M_{\rm body}=10^{23}$ kg $\sim 10M_{\rm Pluto}$ following \cite{Hyodo:2017}, where $M_{\rm Pluto}$ is Pluto mass.
We assume that $10 \%$ of it is captured by Saturn, thus the captured mass is $M_{\rm cap}=10^{22} \ {\rm kg} \sim M_{\rm Pluto}$.
The radius of tidal fragments can be estimated from the condition in which the sum of tensile strength and self-gravity are equal to the tidal force:
\begin{equation}
    -Sr_{\rm p}^2-\frac{Gm_{\rm p}^2}{r_{\rm p}^2}+\frac{2GM_0m_{\rm p}r_{\rm p}}{r^3}=0,
\end{equation}
where $S$, $m_{\rm p}$ and $r_{\rm p}$ are the tensile strength, mass and radius of the fragments.
By solving for $r_{\rm p}$, we can obtain 
\begin{equation}
    r_{\rm p}=\frac{3}{4\pi\rho_{\rm p}}\qty[\frac{S}{G\qty(\frac{3M_0}{2\pi\rho_{\rm p}r^3}-1)}]^{1/2},
\end{equation}
where $\rho_{\rm p}$ is the density of the fragment.
Assuming $\rho_{\rm p}=0.9 \ {\rm g/cm^3}$ and $S=1.6 \ {\rm MPa}$ for ice \citep{Lange:1983, Stewart:1999} and $r=1.1R_0\sim0.5r_{\rm R, Saturn}$, we obtain $r_{\rm p}\sim 200 \ {\rm km}$.
Thus, fragments with a size of $\sim$ a few hundred km are expected to be produced by the tidal disruption of Pluto-sized icy body.
Given the above consideration, in our simulation, the captured fragments are represented as equal-mass particles whose size is $r_{\rm p}=300$ km and density is $\rho_{\rm p}=0.9 \ {\rm g/cm^3}$.
The corresponding fluid Roche limit is $r_{\rm R, Saturn}\simeq 136,000 \ {\rm km}\simeq 2.25R_0$.
The initial number of particles is automatically determined to be $N=98$ from $r_{\rm p}, \rho_{\rm p},$ and $M_{\rm cap}$.
Although tidal fragments are expected to follow a size distribution in reality, the largest fragments, with sizes of a few hundred kilometers, may dominate the mass and thus subsequent dynamical evolution \citep[e.g., Shoemaker-Levy 9; ][]{Weaver:1995}.
Therefore, our simulation results are not expected to change significantly even when an initial size distribution of fragments is included.

\begin{figure}
\centering
\includegraphics[width=\linewidth]{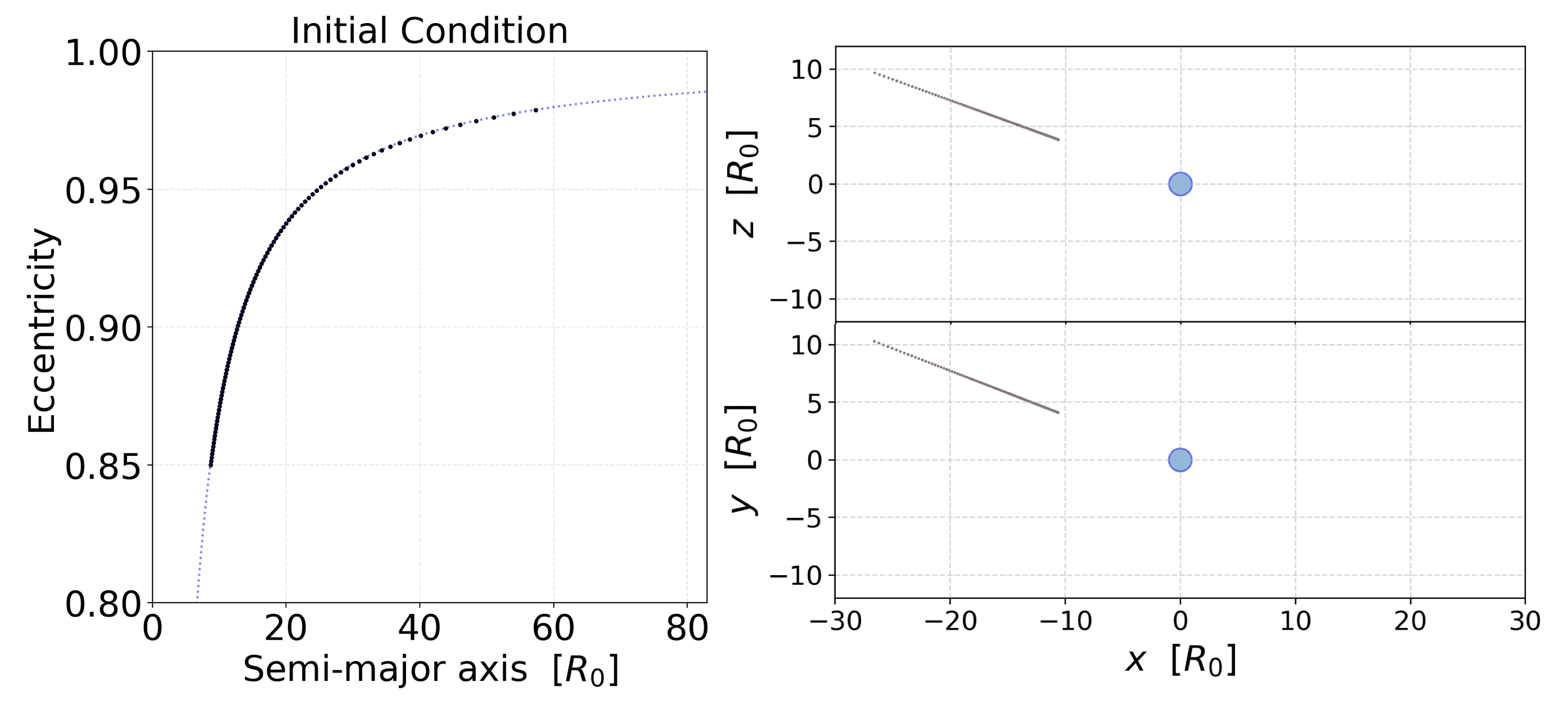}
\caption{The initial condition of our $N$-body simulation in the case of $i_{\rm TD}=20^{\circ}$ and $q_{\rm TD}=1.2R_0$ (Model 2). The left panel shows the $e-a$ distribution. The blue dotted curve represents the analytical distribution of captured fragments just after the tidal disruption by \cite{Hyodo:2017} (Eq. \ref{eq:eaDistribution}). The upper and lower right panels show the snapshots projected on the $xz$ and $xy$ plane, respectively. The blue circle at the center is the central planet and grey dots are initial fragments.}
\label{fig:InitialCondition}
\end{figure}

We set the initial position and velocity of particles so that it mimics the situation just after the tidal disruption, which is explained as follow.
The specific angular momentum of a passing body in a hyperbolic orbit with the pericenter distance to the planet (in this case, Saturn) $q_{\rm TD}$ and the velocity at infinity $v_{\infty}$ is written as \citep{Tremaine:2023}
\begin{align}
     j_{\rm body} &=\sqrt{(q_{\rm TD}v_{\infty})^2 + 2GM_0q_{\rm TD}}\notag\\
     &=\sqrt{q_{\rm TD}^2(v_{\infty}^2 + v_{\rm esc}^2)},\label{eq:AngularMomentum}
\end{align}
where $M_0$ is the mass of the central planet and $v_{\rm esc}$ is the escape velocity at the pericenter distance defined as
\begin{equation}
    v_{\rm esc}=\sqrt{\frac{2GM_0}{q_{\rm TD}}} \label{eq:escape}.
\end{equation}
On the other hand, the specific angular momentum of the captured tidal fragments in a bound orbit with eccentricity $e$ and semi-major axis $a$ is given by $j_{\rm frag}=\sqrt{GM_0a(1-e^2)}$.
By solving for $e$, we obtain the relation
\begin{equation}
    e=\sqrt{1-\frac{j_{\rm frag}^2}{M_0Ga}}.\label{eq:eaDistribution}
\end{equation}
According to \cite{Hyodo:2017}'s SPH simulation (see their Fig. 13), the specific angular momentum of the captured fragments $j_{\rm frag}$ is nearly equal to that of the passing body $j_{\rm body}$.
Thus, we set as an initial condition in this study
\begin{equation}
     j_{\rm frag} = j_{\rm body} =\sqrt{q_{\rm TD}^2(v_{\infty}^2 + v_{\rm esc}^2)}, \label{eq:AngularMomentum}
\end{equation}
which is substituted into Eq. \ref{eq:eaDistribution} (see Fig. \ref{fig:InitialCondition}).
In this paper, we use a fixed velocity of $v_{\infty}=3.0$ km/s, which is the expected value during LHB for Saturn \citep{Charnoz:2009}.
We assign $e$ and $a$ of each particle to follow the relationship of Eq. \eqref{eq:eaDistribution} with the range of $e\in[0.85, 0.98]$.
Their initial inclination $i$ are set to be equal to that of the orbit of the tidally disrupted body $i_{\rm TD}$.
The argument of pericenter and the ascending node of all particles are assumed to be aligned.
Their initial true anomaly $f$ is set to be $f=160^{\circ}$.

In the following sections, we vary the value of the inclination $i_{\rm TD}$ and pericenter distance $q_{\rm TD}$ of the tidal disruption and investigate these dependence on the following dynamical evolution.
The parameter sets of our simulation is summarized in Table \ref{tab:run_summary}.
Figure \ref{fig:InitialCondition} shows the initial condition of our simulations in the case of $i_{\rm TD}=20^{\circ}$ and $q_{\rm TD}=1.2R_0$ (Model 2) as an example.
The left panel shows the $e-a$ distribution of particles and the right panels show the snapshots projected on $x-z$ (upper panel) and $x-y$ (lower panel) planes.
The blue dotted curve represents the analytical estimate of $e-a$ distribution of captured fragments (Eq. \ref{eq:eaDistribution}).
All the particles strictly lie on the blue dotted curve in the initial condition.

\begin{table*}[h]
 \caption{The summary of parameters in our $N$-body simulations. $i_{\rm TD}$ and $q_{\rm TD}$ are the inclination and the periceter distance of the tidal disruption, respectively.}
 \centering
  \begin{tabular}{ccc}
   \hline
      & $i_{\rm TD}$  & $q_{\rm TD}$  [$R_0$] \\
   \hline \hline
   Model 1 & $1.0^{\circ}$ & $1.2$ \\
   Model 2 & $20^{\circ}$ & $1.2$  \\
   Model 3 & $35^{\circ}$ & $1.1$  \\
   Model 4 & $45^{\circ}$ & $1.1$  \\
   Model 5 & $60^{\circ}$ & $1.1$  \\
   \hline
   \label{tab:run_summary}
  \end{tabular}
\end{table*}

\section{Analytical Arguments}\label{sec:Analytical Argument}
In this section, we summarize the analytical arguments necessary to interpret our $N$-body simulation results shown in the following section.
\subsection{Precession rates}\label{sec:PrecessionRate}
The argument of pericenter $\omega$ and the longitude of ascending node $\Omega$ of orbits around flattened body precess due to the non-spherical $J_2$ term \citep{Murray:1999}.
Their precession rates are given by \citep{Kaula:1966}
\begin{align}
    \dot{\omega}&=\frac{3nJ_2}{(1-e^2)^2}\qty(\frac{R_0}{a})^2\qty(1-\frac{5}{4}\sin^2i), \label{eq:omegaPrecession}\\
    \dot{\Omega}&=\frac{3nJ_2}{2(1-e^2)^2}\qty(\frac{R_0}{a})^2\cos i \label{eq:OmegaPrecession},
\end{align}
where $n$ is the orbital mean motion.
Thus, the precession rate of longitude of pericenter $\varpi=\omega+\Omega$ is 
\begin{align}
    \dot{\varpi}&=\dot{\omega}+\dot{\Omega}\\
    &=\frac{3nJ_2}{(1-e^2)^2}\qty(\frac{R_0}{a})^2\qty(1-\frac{5}{4}\sin^2 i-\frac
{1}{2}\cos i). \label{eq:piPrecession}
\end{align}

\begin{figure}
\centering
\includegraphics[width=0.7\linewidth]{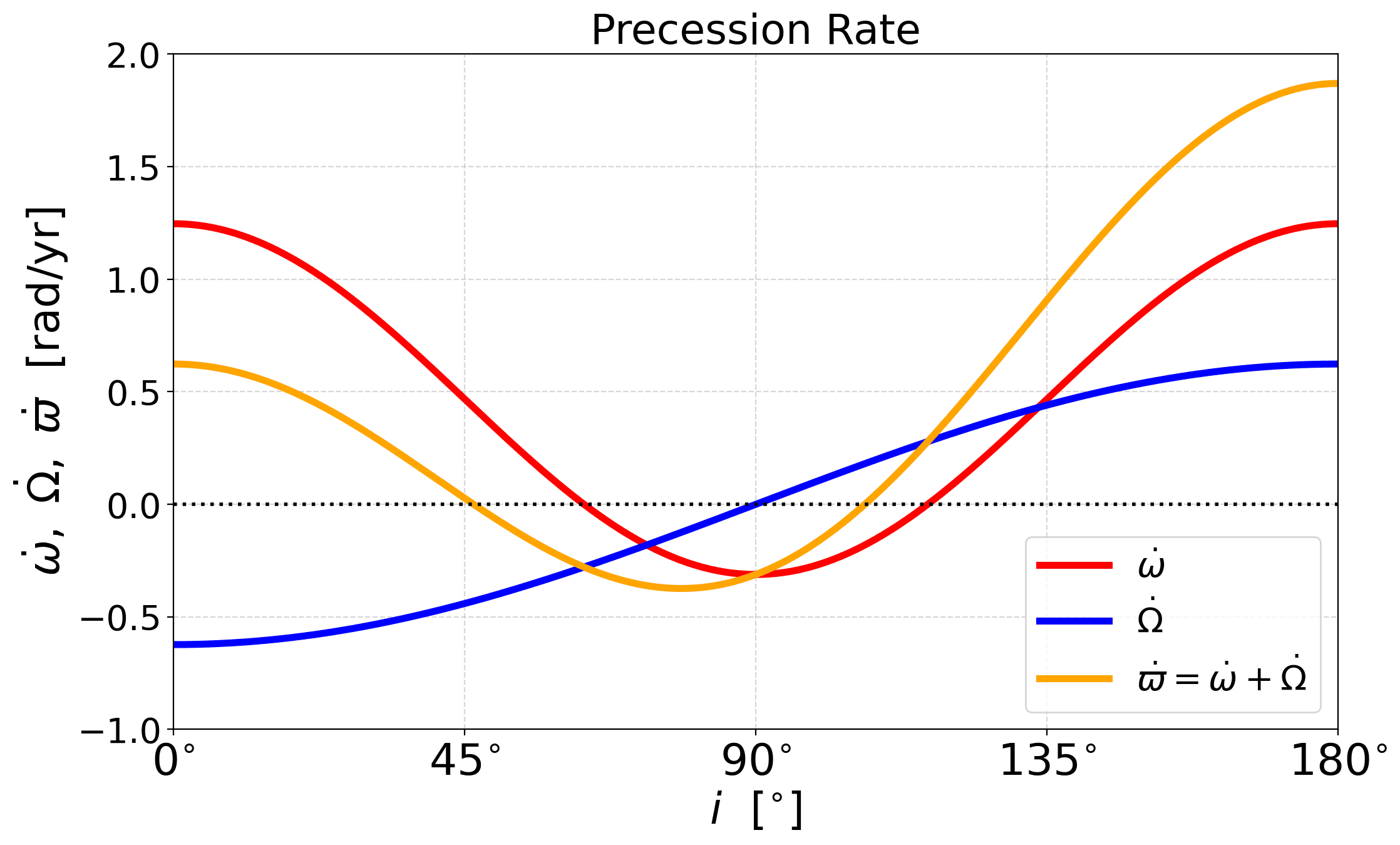}
\caption{The precession rates of $\omega$ (red), $\Omega$ (blue) and $\varpi$ (orange) in the case of $a=20R_0$ and eccentricity calculated by Eq. \ref{eq:eaDistribution}.}
\label{fig:Precession Rate}
\end{figure}

Figure \ref{fig:Precession Rate} shows the precession rates as a function of inclination in the case of $a=20R_0$ and $e \sim 0.938$ which is calculated by Eq. \ref{eq:eaDistribution}.
We find that there are two critical inclinations where the $\dot{\omega}$ and $\dot{\Omega}$ are canceled out ($\dot{\varpi}=0$).
If we equate the right hand side of Eq. \eqref{eq:piPrecession} to zero, we obtain the critical inclination $i_{c+}$ and $i_{c-}$:
\begin{align}
    i_{c+}&=\arccos\left(\frac{1+\sqrt{6}}{5}\right)\sim 46.378^{\circ},\label{eq:ic+} \\
    i_{c-}&=\arccos\left(\frac{1-\sqrt{6}}{5}\right)\sim 106.852^{\circ}.\label{eq:ic-}
\end{align}
Note that this critical value does not depend on the physical parameters of central planets, such as $J_2$ and $R_0$, thus this analysis can be directly applied to any planets with $J_2$ term.

\cite{Hyodo:2017} analytically predicted that, comparing $\tau_{\omega}=2\pi/\dot{\omega}$ and $\tau_{\Omega}=2\pi/\dot{\Omega}$ with collisional timescale $\tau_{\rm col}$, the orbital precession occurs faster than the onset of collisional grinding, as a result, a torus-like structure firstly forms by the differential precession, following the collisional grinding.
Here, we argue that the timescale of torus-like structure formation should be evaluated with the precession timescale of longitude of pericenter $\tau_{\varpi}=2\pi/\dot{\varpi}$.
It becomes longer when the inclination is close to $i_{c+}$ or $i_{c-}$, thus significant collisional grinding may occur before forming the torus-like structure.
As a results, the dynamical path should depend on the inclination and the dynamical picture predicted by \cite{Hyodo:2017} could be modified (see the discussion in Section~\ref{sec:EqRadius}).
We demonstrate this argument through our $N$-body simulation including collisional grinding focusing on the prograde cases in Section \ref{sec:Simulation Results}.

\subsection{Equivalent circular orbital radius}\label{sec:EqRadius}
In this subsection, we discuss the ring structure which finally forms as a result of collisional evolution.
In the previous literature, the equivalent circular orbital radius are commonly used for this analysis, but they did not consider the effect of inclination \citep[e.g.,][]{Hyodo:2017, Kegerreis:2025}.
Here, we revisit the derivation of equivalent circular radius with considering the inclination and show that we can obtain a constraint on the inclination so that enough material finally survives to form the ring-satellite system.

The gravitational potential including $J_2$ term is axisymmetric (see Eq. \eqref{eq:gravitational potential}).
Therefore, the $z$ component of total angular momentum of the system is conserved during the dynamical evolution.
The $z$ component of specific angular momentum of a fragment just after the tidal disruption is $j_{\rm frag, z}=j_{\rm frag}\cos i_{\rm TD}$.
When $e$ and $i$ of the particle damp due to collisions between other particles conserving $j_{\rm frag, z}$, its orbital radius of circular ($e=0$) and equatorial ($i=0)$ orbit which is finally established is given by
\begin{align}
    a_{\rm eq} &= \frac{j_{\rm frag}^2}{GM_0}\cos^2 i_{\rm TD}\\
    &= 2q_{\rm TD}\qty(\frac{v_{\infty}^2}{v_{\rm esc}^2} + 1)\cos^2 i_{\rm TD},  \label{eq:a_eq}
\end{align}
where we use the relation Eq. \eqref{eq:AngularMomentum}.
The escape velocity at the pericenter distance $v_{\rm esc}$ of Jupiter, Saturn, Uranus and Neptune are $\sim 54, \ 32, \ 19$ and $21 \ {\rm km/s}$, respectively, where we consider the case of $q_{\rm TD}=1.2R_0$.
These are about 10 times larger than $v_{\infty}$ within the realistic range of $0.0 \ {\rm km/s} \leq v_{\infty}\leq3.0 \ {\rm km/s}$.
The first term in the parenthesis of Eq. \eqref{eq:a_eq} is much smaller than unity and changing $v_{\infty}$ in this range, thus changing the central planet has almost no effect on the value of $a_{\rm eq}$.

\begin{figure}
\centering
\includegraphics[width=\linewidth]{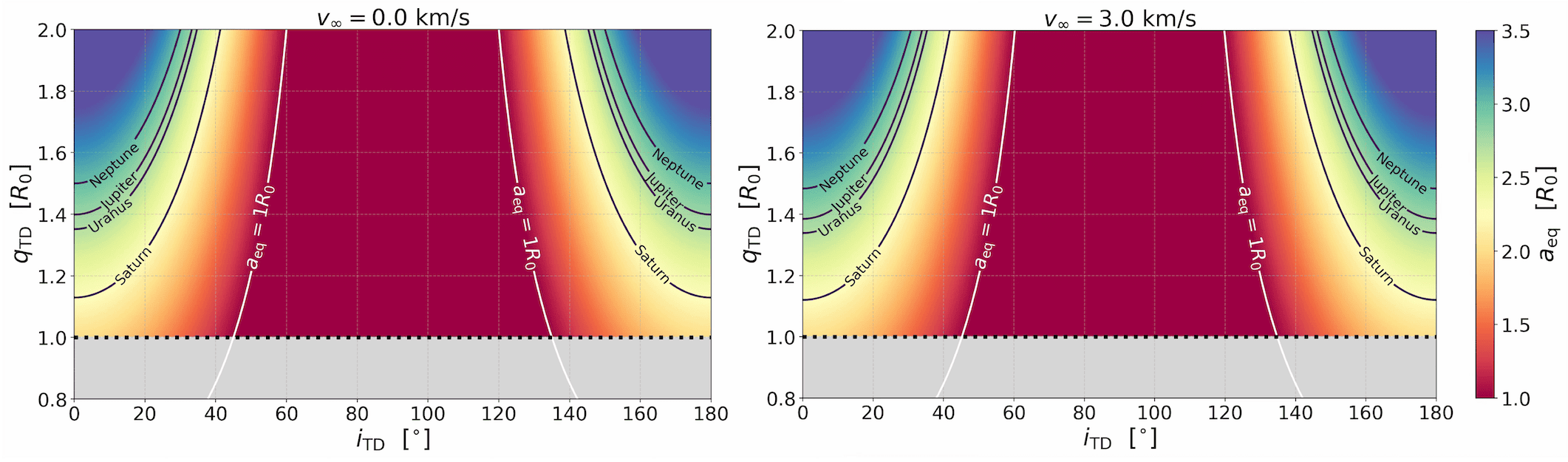}
\caption{The equivalent orbital radius is shown by color map on the $(q_{\rm TD}, i_{\rm TD})$ plane for all giant planets in the Solar System. The curve corresponding to $a_{\rm eq}=1R_0$ are shown in white. The curves corresponding to $a_{\rm eq}=r_{\rm R, Planet}$ are also shown and labeled with the name of each planet, where $r_{\rm R, Planet}$ are the planet's Roche limit radius for icy material. 
There is little dependence of $v_{\infty}$, which is discussed in the main text.}
\label{fig:qi-space}
\end{figure}

Because $a_{\rm eq}$ depends only on $q_{\rm TD}$ and $i_{\rm TD}$ (see Eq. \eqref{eq:a_eq}) and little dependence of changing the central planet on $a_{\rm eq}$, we can draw it in the $(q_{\rm TD}, i_{\rm TD})$ parameter space for all giant planets in the Solar System as shown in Fig. \ref{fig:qi-space}.
To show the little dependence of $v_{\infty}$, we plot the two cases with $v_{\infty}=0$ and $3 \ {\rm km/s}$.
Because the fluid Roche limit depends on the planet density (see Eq. \ref{eq:Roche}),  we also plot the curves corresponding to $a_{\rm eq}=r_{\rm R, Planet}$ and $a_{\rm eq}=1R_0$ for different giant planets, where $r_{\rm R, Planet}$ is the classical fluid Roche limit radius of the planet for icy material, which is defined as 
\begin{equation}
    r_{\rm R, Planet} = 2.456 \left(\frac{\rho_0}{\rho_{\rm p}} \right)^{1/3} R_0,\label{eq:Roche}
\end{equation}
where $\rho_0$ is the planet bulk density, $\rho_{\rm p}=0.9 \ {\rm g/cm^3}$ is the density of ice and $R_0$ is the planet radius.
The each curve is labeled with the name of the planet.
The Roche limit radius normalized with the planet radius is smaller for Saturn than that for the other planets because of the Saturn's smallest bulk density.
Comparing with the each panel, there is little dependence of $v_{\infty}$ due to the reason desrcibed above.

These figures show that the orbital radius of the finally formed rings becomes smaller with approaching $i_{\rm TD}=90^{\circ}$.
We find that it ranges from $<1R_0$ to close to or beyond $r_{\rm R, ice}$ depending on $i_{\rm TD}$ and $q_{\rm TD}$.
More importantly, if $a_{\rm eq}$ is very close to or smaller than $R_0$, we expect that significant amount of material would fall onto the planet, which decreases the mass delivered to the finally formed ring-satellite system.
On the other hand, if the pericenter distance of tidal disruption becomes large, the mass of captured fragments itself significantly decrease.
For example, the boundary where the capture efficiency becomes smaller than $\sim 1\%$ is $\sim 1.4R_0$ and $\sim 1.8R_0$ in the case of Saturn and Uranus, respectively \citep[][]{Hyodo:2017}.

Previous studies \citep[e.g., ][]{Crida:2012} proposed that the regular satellites around giant planets in the Solar System formed from a primordial massive ring.
However, this scenario requires a primordial ring mass at least including the current rings and ring-originated satellites.
Thus, from the above analytical prediction and previous SPH simulations \citep{Hyodo:2017}, we can constrain the required values of $(q_{\rm TD}, i_{\rm TD})$ for the tidal disruption to supply enough mass, assuming that massive rings followed by the ring-satellite system formation were originated from a single tidal disruption event.
In the following sections, we demonstrate this prediction through our direct $N$-body simulations considering Saturnian system as an example.

We summarize $\tau_{\varpi}$ and $a_{\rm eq}$ values in our models calculated with Eq. \eqref{eq:piPrecession} and Eq. \eqref{eq:a_eq}, respectively.
When calculating $\tau_{\varpi}$, we use $a=10R_0$ and the corresponding eccentricity calculated by Eq. \eqref{eq:eaDistribution} as a reference.
The precession timescale in Model 4 is much longer than the case of other models because $i_{\rm TD}$ is close to $i_{\rm c}$, thus the collisional grinding could be triggered before the formation of a torus-like structure (for more quantitative discussion, see Section \ref{sec:DynamicalPath}).
The value of $a_{\rm eq}$ in Model 1, 2 and 3 is larger than the planet radius, thus it is expected that most of fragments can survive and massive ring can form.
On the other hand, $a_{\rm eq}$ in Model 4 is very close to the planet radius while $a_{\rm eq}$ in Model 5 is smaller than it, thus most of fragments are expected to fall onto the planet as a result of $e$ and $i$ damping.

\begin{table*}[h]
 \caption{The summary of calculated $\tau_{\omega}$, $\tau_{\Omega}$, $\tau_{\varpi}$ and $a_{\rm eq}$ in our each model.  We consider $a=10R_0$ and eccentricity calculated by Eq. \ref{eq:eaDistribution} as a reference. 
 The plus/minus of the timescale means that the precession is prograde/retrograde. 
 The value of $a_{\rm eq}$ in the parenthesis is in the unit of the Roche limit of Saturn.}
 \centering
  \begin{tabular}{ccccccc}
   \hline
      & $\tau_{\omega}$ [yr] & $\tau_{\Omega}$ [yr] & $\tau_{\varpi}$ [yr] & $a_{\rm eq}$  [$R_0$ ($r_{\rm R}$)] \\
   \hline \hline
   Model 1 & 1.783 & -3.565 & 3.567 & 2.40 (1.06) \\
   Model 2 & 2.087 & -3.793 & 4.642 & 2.12 (0.94) \\
   Model 3 & 2.543 & -3.656 & 8.357 & 1.48 (0.65)  \\
   Model 4 & 3.993 & -4.235 & 69.823 & 1.10 (0.49) \\
   Model 5 & 23.959 & -5.990 & -7.986 &  0.55 (0.24)\\
   
   \hline
   \label{tab:value_summary}
  \end{tabular}
\end{table*}

\section{Simulation Results}\label{sec:Simulation Results}
In this section, we present our results of $N$-body simulations. 
First, we show the overview of the dynamical path in the case of  $(q_{\rm TD}, i_{\rm TD})=(1.2R_0, 1.0^{\circ})$ and $(1.2R_0, 20^{\circ})$, and discuss the effect of inclination in subsection \ref{sec:DynamicalPath}.
Then, we discuss the structure of the circularized ring which eventually forms as a result of collisional evolution comparing the case of $(q_{\rm TD}, i_{\rm TD})=(1.2R_0, 20^{\circ})$ and $(1.1R_0, 35^{\circ})$ in subsection \ref{sec:NarrowRings}.
We demonstrate that the prediction described in Section \ref{sec:EqRadius} by showing the results with $(q_{\rm TD}, i_{\rm TD})=(1.1R_0, 45^{\circ})$ and $(1.1R_0, 60^{\circ})$ and discuss the constraint of $(q_{\rm TD}, i_{\rm TD})$ parameter space in subsection \ref{sec:larger_i}.
Finally, we compile our simulation results and discuss the classification of the dynamical path and fate on $(q_{\rm TD}, i_{\rm TD})$ parameter space in subsection \ref{sec:constraint}.

\subsection{The dependence of the inclination on dynamical path}\label{sec:DynamicalPath}

\begin{figure}
\centering
\includegraphics[width=\linewidth]{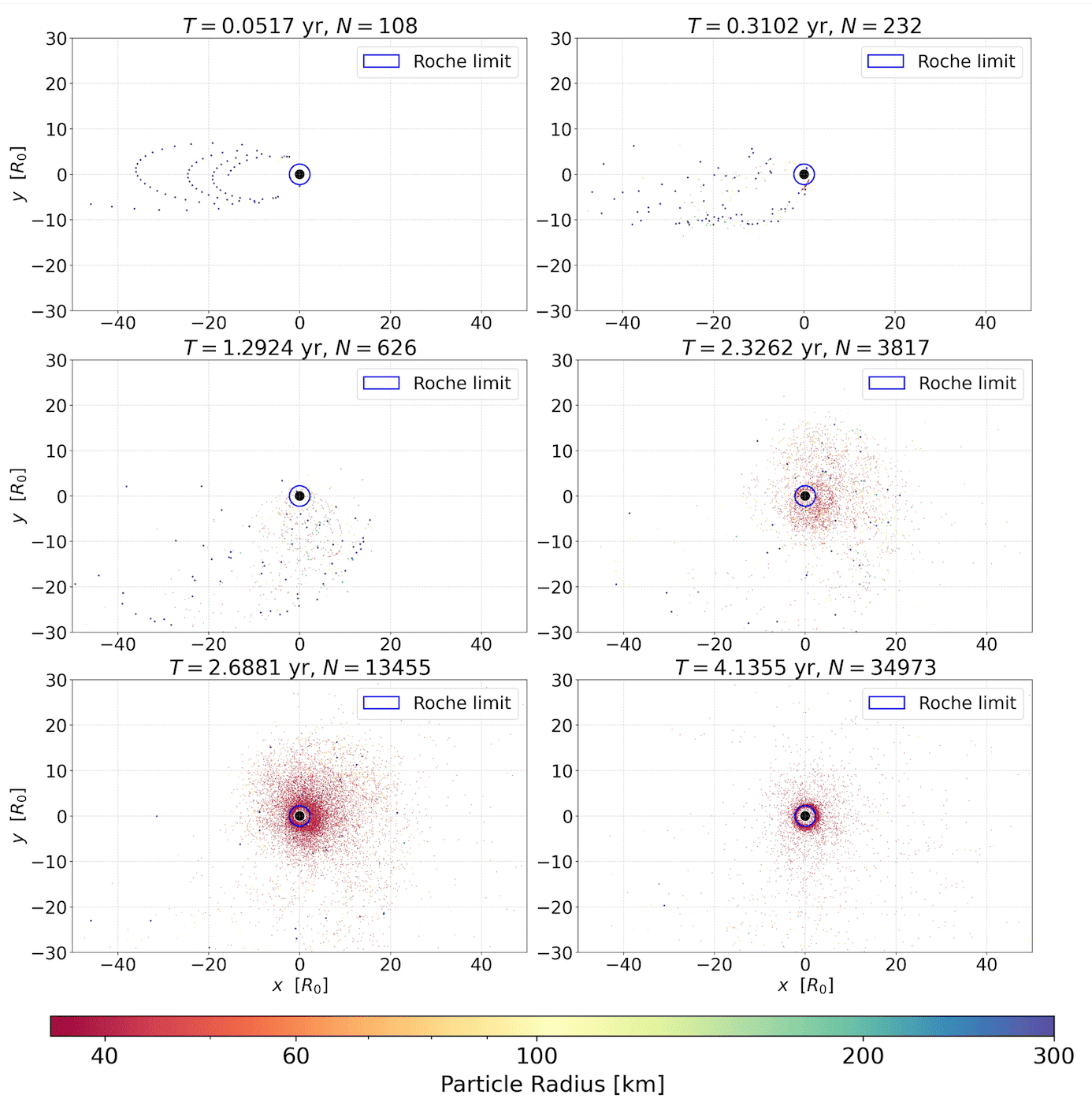}
\caption{The time evolution of simulated collisional system in the case of $i_{\rm TD}=1.0^{\circ}$ and $q_{\rm TD}=1.2 R_0$ (Model 1). The color of each plot shows the particle radius. The time and the number of particles are shown on the top of each panel.}
\label{fig:snap_i1.0}
\end{figure}

\begin{figure}
\centering
\includegraphics[width=\linewidth]{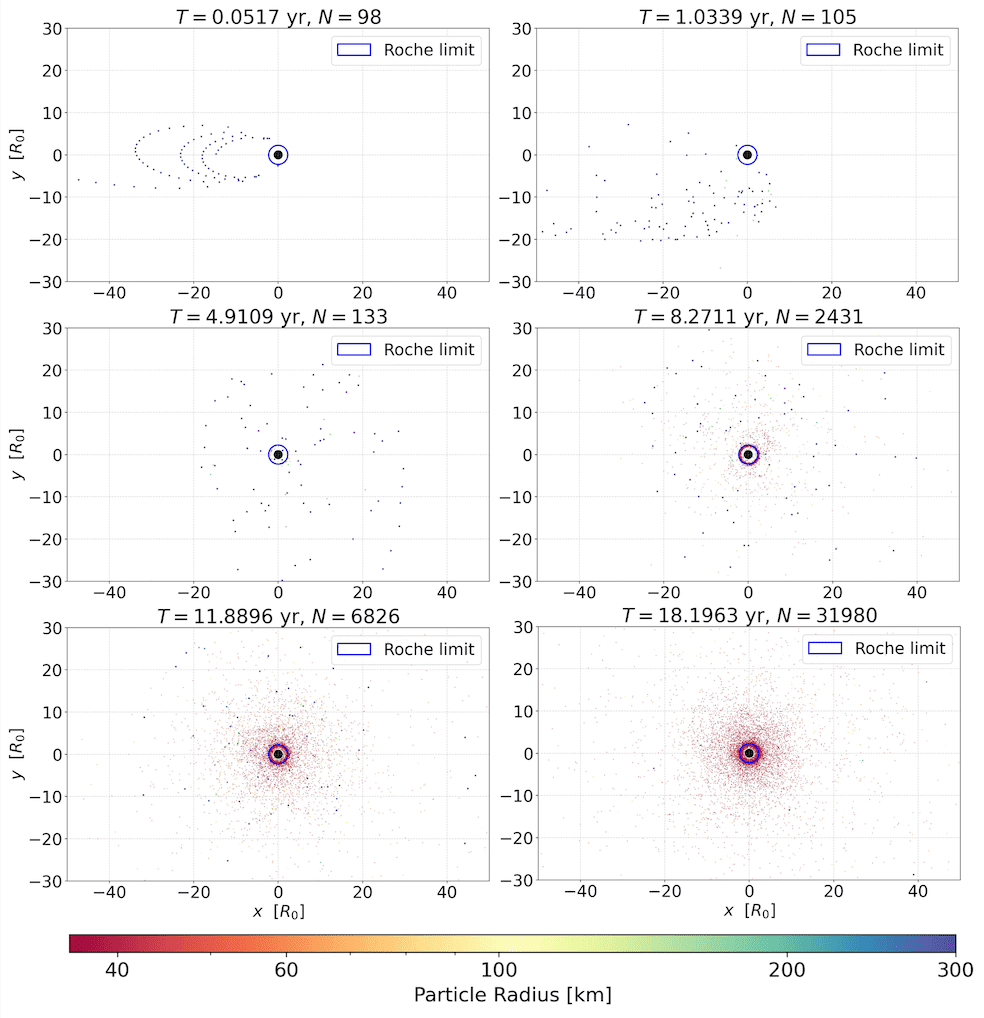}
\caption{Same as Fig. \ref{fig:snap_i1.0}, but for $i_{\rm TD}=20^{\circ}$ and $q_{\rm TD}=1.2 R_0$ (Model 2).}
\label{fig:snap_i20}
\end{figure}

\begin{figure}
\centering
\includegraphics[width=\linewidth]{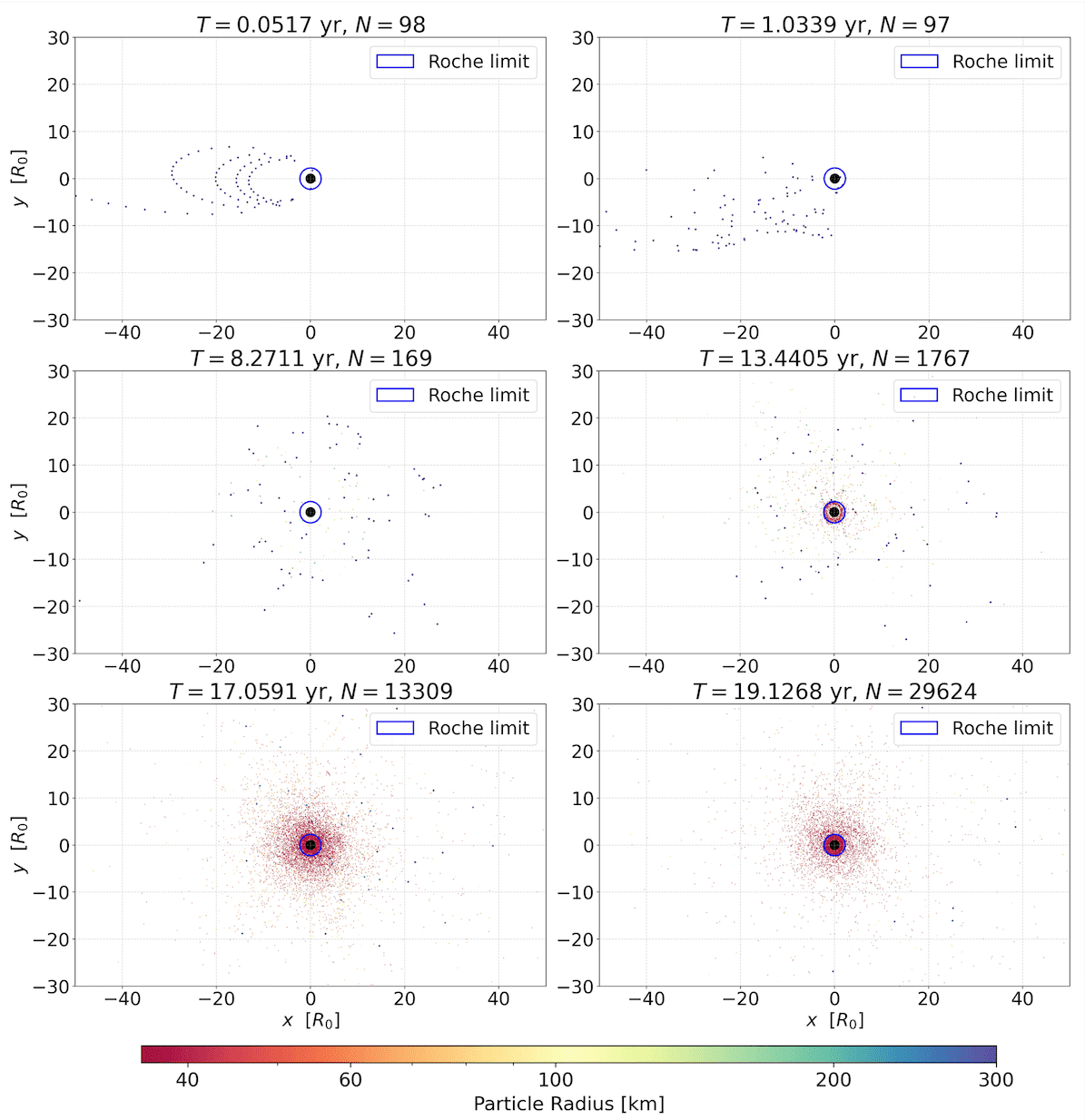}
\caption{Same as Fig. \ref{fig:snap_i1.0}, but for $i_{\rm TD}=35^{\circ}$ and $q_{\rm TD}=1.1 R_0$ (Model 3).}
\label{fig:snap_i35}
\end{figure}

Figures~\ref{fig:snap_i1.0}, \ref{fig:snap_i20} and \ref{fig:snap_i35} show the snapshots of the time evolution in Model 1, 2 and 3, respectively.
The blue circle at the origin represents Saturn and dotted circle shows its Roche limit radius.
The plotted size of each particle reflect its physical radius.
The color represents their radius in Figs. \ref{fig:snap_i1.0}, \ref{fig:snap_i20} and \ref{fig:snap_i35}.
In Appendix A, we also present the zoom-in snapshots of each model from top and side view in Figs. \ref{fig:snap_i1.0_zoom}, \ref{fig:snap_i20_zoom}, \ref{fig:snap_i35_zoom}, \ref{fig:snap_i1.0_zoom_z}, \ref{fig:snap_i20_zoom_z} and \ref{fig:snap_i35_zoom_z}.
It is clear that, in Model 1 ($i_{\rm TD}=1.0^{\circ}$), the significant collisional grinding is triggered before $\varpi$ is fully randomized by differential precession.
After the onset of collisional grinding, the number of particles shown on the top of the each panel rapidly increases, and a circular ring eventually forms at the orbital radius $\sim r_{\rm R}$ (see Fig.~\ref{fig:snap_i1.0_zoom}), which is consistent with the analytically calculated equivalent orbital radius $a_{\rm eq}\sim 1.06 \, r_{\rm R}$ (Table \ref{tab:value_summary}).
The orbitally-circularized particles are radially accumulated around $a_{\rm eq}$ and the width of the rings is narrow, which is discussed in Section \ref{sec:NarrowRings} in detail.
On the other hand, in Model 2 ($i_{\rm TD}=20^{\circ}$) and Model 3 ($i_{\rm TD}=35^{\circ}$), significant collisional grinding does not occur until $\sim \tau_{\varpi}$.
This is because the collision probability during the precession is lower than the case of $i_{\rm TD}=1.0^{\circ}$ due to the larger spatial dispersion of the system by its larger inclination (see also Fig. \ref{fig:snap_i20_zoom_z} and Fig. \ref{fig:snap_i35_zoom_z}).
As a result, a torus-like structure firstly forms, then the collisional grinding is triggered, which is the consistent dynamical picture with that described by \cite{Hyodo:2017}.
The orbital radius of finally-formed circular ring is close to the Roche limit in Model 2 (see Fig, \ref{fig:snap_i20_zoom}), while it is between the Saturn's surface and its Roche limit in Model 3, which is consistent with the calculated equivalent orbital radius $a_{\rm eq}\sim 0.94 \, r_{\rm R}$ for Model 2 and $a_{\rm eq}\sim 0.65 \, r_{\rm R}$ for Model 3 (Table \ref{tab:value_summary}).

\begin{figure}
\centering
\includegraphics[width=0.7\linewidth]{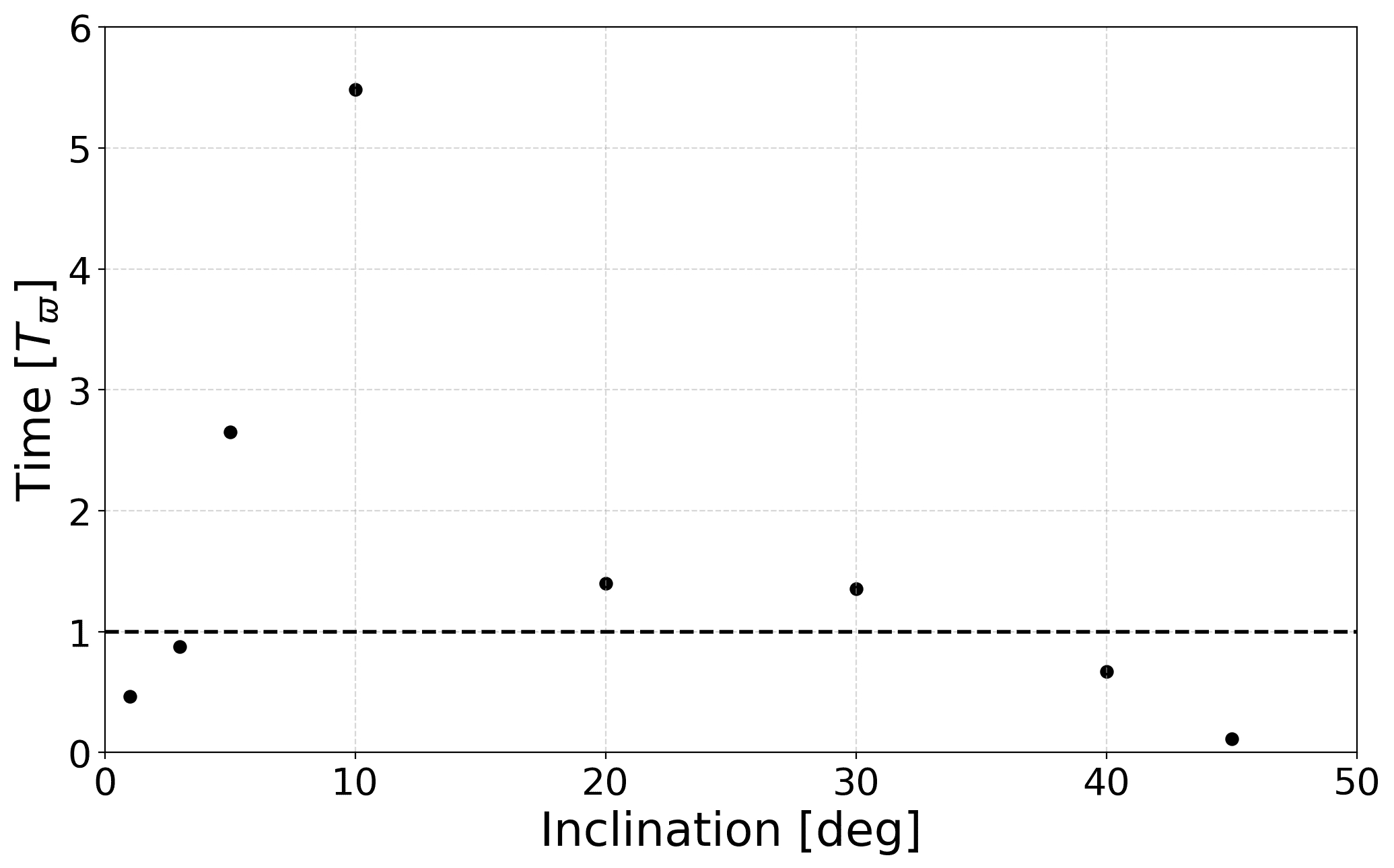}
\caption{The time when the collisional grinding is triggered in the unit of $\varpi$ precession time in the additional runs with several $i_{\rm TD}$.
We define that the collisional grinding is triggered when the number of particles exceeds 1000.}
\label{fig:Tcol=Tpre}
\end{figure}

In order to quantify the inclination range where the \cite{Hyodo:2017}'s timescale comparison is affected, we additionally simulate the case with several $i_{\rm TD}$ values in the range between $0^{\circ}$ and $45^{\circ}$ with $q_{\rm TD}=1.2R_0$.
Figure \ref{fig:Tcol=Tpre} shows the time when the collisional grinding is triggered in each additional run in the unit of $\varpi$ precession timescale $T_{\varpi}$, where we define that the collisional grinding is triggered when the number of particles exceeds $1000$ ($\sim10$ times the initial value) \footnote{
This definition is somewhat arbitrary, but changing this threshold value does not affect our conclusion here.
}.
The dashed line shows the case where the onset of collisional grinding coincides with the completion of $\varpi$ precession.
When $i_{\rm TD}$ is very small ($i_{\rm TD}\lesssim 3.0^{\circ}$) or $i_{\rm TD} \gtrsim 35^{\circ}$ (close to $i_{\rm c} \simeq 46.378^{\circ}$), the collisional grinding begins before the completion of $\varpi$ precession, which modifies the dynamical path in \cite{Hyodo:2017}.
On the other hand, when $i_{\rm TD}$ is moderate value ($3.0^{\circ}\lesssim i_{\rm TD} \lesssim 35^{\circ}$), the $\varpi$ precession is firstly completed followed by the onset of collisional grinding.

Figure \ref{fig:eccDapming_Model1,2} shows the time evolution of the mass of particles with eccentricity less than 0.2, $M_{e<0.2}$, in Model 1 and 2, and Figure \ref{fig:incDapming_Model2} shows the time evolution of that of particles with inclination less than $5.0^{\circ}$, $M_{i<5.0^{\circ}}$, in Model 2. 
As a reference, we show the present Saturn's ring mass evaluated by \cite{Iess:2019} with the red dotted line and the mass including that of the present rings and all moons orbiting inside Titan with the blue dotted line, which was proposed to be formed from a spreading primordial massive rings by \cite{Crida:2012}.
As mentioned in the above, the increase of the mass corresponding to the onset of collisional grinding is earlier in Model 1 than Model 2 because of its higher collision probability during the precession.
In the case of Model 1, $M_{e<0.2}$ increases and exceeds the red and blue lines within $\sim 5$ years, which means that the circular ring can form in a few years in this case.
Its mass would be large enough to supply the present rings and inner satellites \citep{Crida:2012}.
The collision velocity is relatively low (see Fig. \ref{fig:vimp}) when the collisional grinding is triggered, because the argument of pericenter is relatively aligned at that time.
As a result, the eccentricity damping by a single collision is moderate and the increase of $M_{e<0.2}$ is relatively gradual.
On the other hand, in the case of Model 2, the argument of pericenter has already randomized when the collisional grinding is triggered.
Thus, higher-velocity collisions occur between particles and the eccentricity and inclination damping by a single collision is more significant.
As a result, $M_{e<0.2}$ and $M_{i<5.0^{\circ}}$ exceed the present ring mass by a few collisions, thus its abrupt increase occurs at $\sim 5$ years.
After that, the collisional grinding proceeds, and circular and equatorial ring with mass larger than that of the present rings and inner moons eventually formed as in the case of Model 1 (see also the $x-z$ plots in Fig. \ref{fig:snap_i1.0_zoom_z} and \ref{fig:snap_i20_zoom_z}). 
\begin{figure}
\centering
\includegraphics[width=\linewidth]{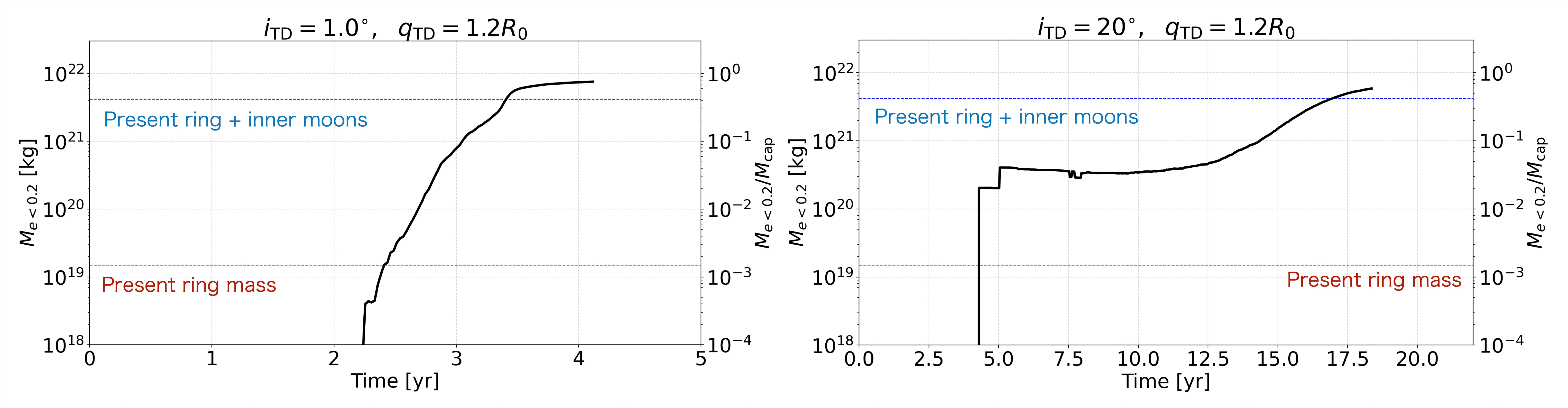}
\caption{The time evolution of the mass with $e<0.2$ in Model 1 (left panel) and 2 (right panel). The red and blue dotted lines represent the mass of present rings mass \citep{Iess:2019} and the mass including the present rings and the inner satellite orbiting inside Titan. The left axis is in the unit of kg and the right axis is normarized by the captured mass $M_{\rm cap}$.}
\label{fig:eccDapming_Model1,2}
\end{figure}

\begin{figure}
\centering
\includegraphics[width=0.7\linewidth]{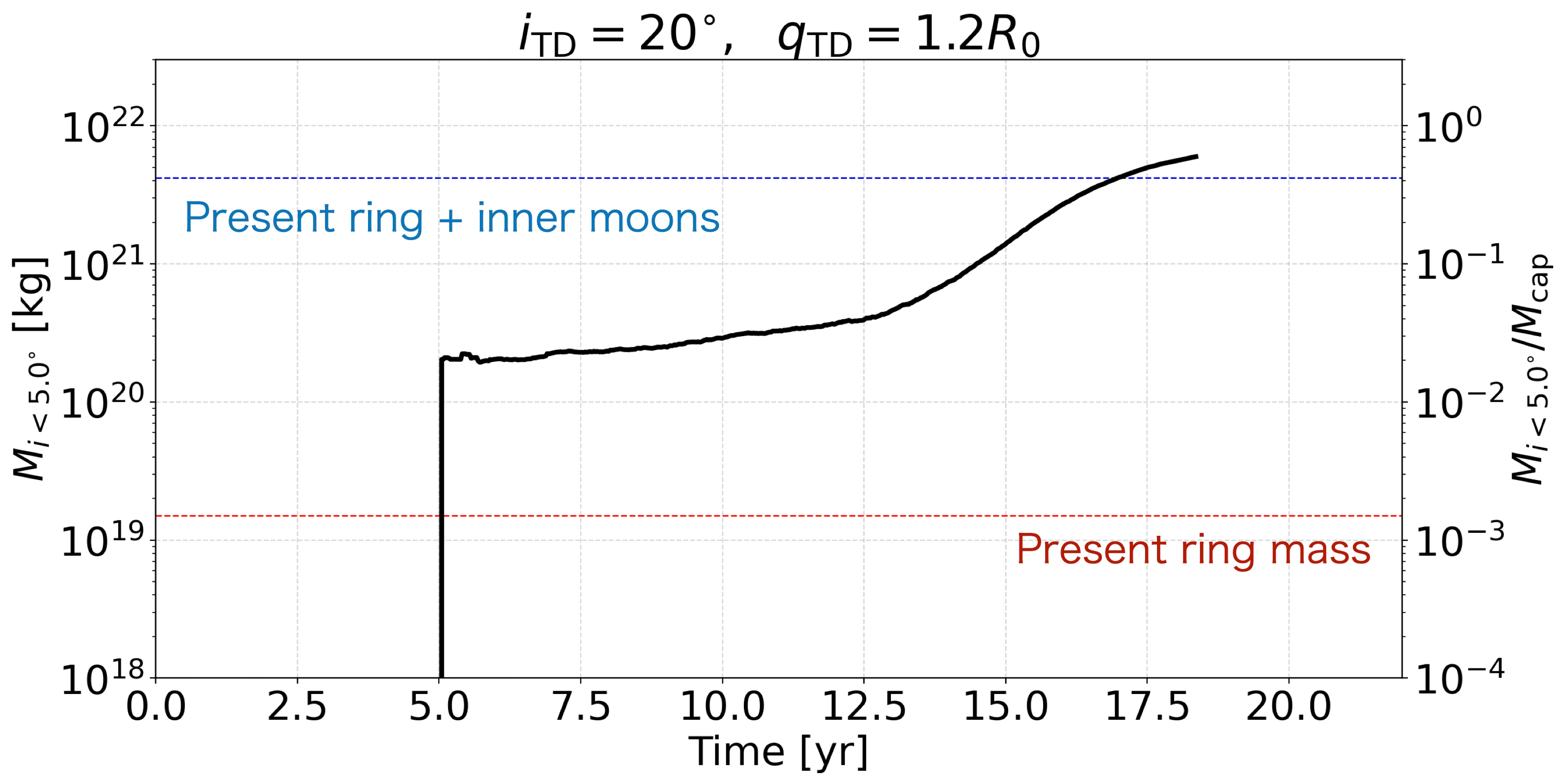}
\caption{The time evolution of the mass with $i<5^{\circ}$ in Model 2. The representation of the axis and dotted lines for reference are same as Fig. \ref{fig:eccDapming_Model1,2}.}
\label{fig:incDapming_Model2}
\end{figure}

As shown the above, we find that the dynamical path, especially the timing of the onset of significant collisional grinding, depends on the inclination.
The onset time of the collisional grinding is earlier when the system is almost on the equatorial plane ($i_{\rm TD}=1.0^{\circ}$) because the system is nearly on the 2D plane, while it is much later than the $\varpi$ precession timescale when the system is spatially more dispersed ($i_{\rm TD}=20^{\circ}$).
Nevertheless, we find that, in both cases, the circular and equatorial ring finally forms as a result of collisional damping of $e$ and $i$.
In addition, the validity of our analytical estimate of $a_{\rm eq}$ is also confirmed by our simulations.
The mass of the circularized equatorial ring is large enough to form the current rings and inner satellites of Saturn if we consider a single tidal disruption of a body of $\sim 10M_{\rm Pluto}$ mass with the inclination and pericenter distance we examined here.
In the next section, we discuss the detail structure of the formed narrow rings and present the implications for the ring and satellite formation from the massive rings in Section \ref{sec:Implication_narrow}.

\subsection{The formation of narrow rings}\label{sec:NarrowRings}
\begin{figure}
\centering
\includegraphics[width=0.8\linewidth]{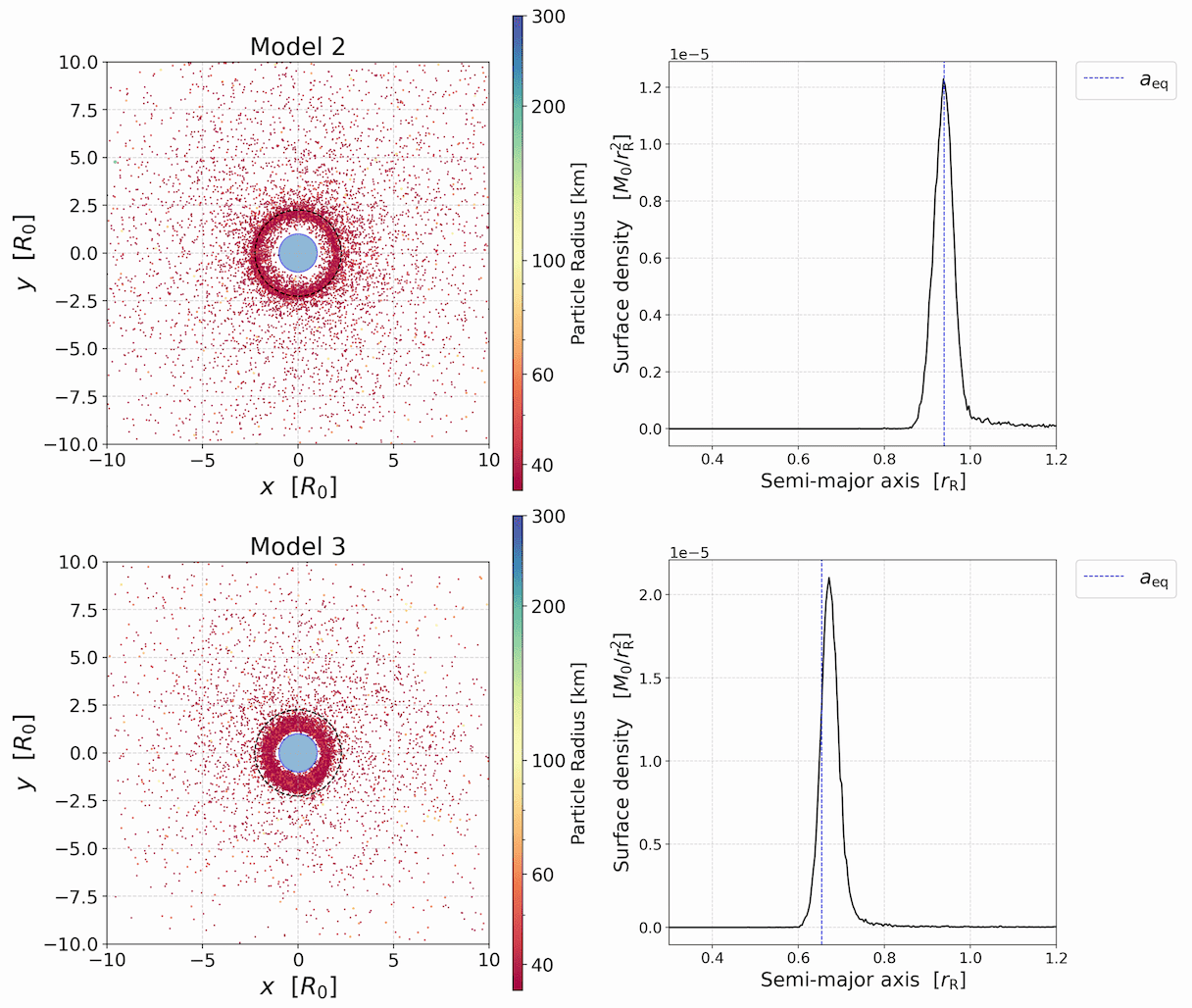}
\caption{The snapshot of the narrow ring (left panel) and its surface density (right panel). The color of particles in the left panel shows the particle radius. The blue dotted line in the right panel shows the equivalent radius calculated with Eq. \ref{eq:a_eq}.}
\label{fig:SurfaceDensity}
\end{figure}

To highlight the structure of finally circularized ring, we show their snapshots and surface density in Fig. \ref{fig:SurfaceDensity}.
The black dashed circle represents the Saturn's Roche limit radius.
The equivalent radius in Model 2 and 3 are $a_{\rm eq}\sim 0.9 \,r_{\rm R}$ and $a_{\rm eq}\sim 0.6 \,r_{\rm R}$, respectively, which is shown as the blue dotted lines in the surface density plots.
It is clear that the surface density has a single peak at $\sim a_{\rm eq}$ and narrow ring is created at the orbital radius $\sim a_{\rm eq}$, which is consistent with the analytical prediction described in Section \ref{sec:EqRadius}
\footnote{
As noted in Section \ref{sec:settings}, our simulation does not include the viscous effects due to the inelastic collisions and self-gravity wakes.
We show that the viscous timescale of the narrow rings is much larger than its formation timescale resulting from the damping of eccentricity and inclination in Appendix B.
Therefore, we argue that our collision outcome treatment without the effect of ring viscosity is valid at least in our simulated timescale to form the narrow ring.
}.
These results indicate that tidal disruption event leads to the formation of a narrow ring as a result of the subsequent collisional evolution due to the conservation of $z$ component of the system's angular momentum. After that, we expect that the narrow rings would viscously diffuse in a longer timescale.

The orbital radius of the created narrow ring is fully determined by the inclination $i_{\rm TD}$ and pericenter distance $q_{\rm TD}$ of the tidal disruption, thus the rings with wide range of the orbital radius from $\sim R_0$ to $\gtrsim r_{\rm R}$ can be formed depending on the configuration of tidal disruption.
The implication from these results on primordial evolution of the ring-satellite system will be discussed in Section \ref{sec:Implication_narrow}.

\subsection{Mass loss of fragments during the $e$ and $i$ damping} \label{sec:larger_i}
Here, we present the results of Model 4 with $(q_{\rm TD}, i_{\rm TD})=(1.1R_0, 45^{\circ})$ and Model 5 with $(q_{\rm TD}, i_{\rm TD})=(1.1R_0, 60^{\circ})$.
In Model 4, the $\varpi$ precession timescale is much longer than in the cases of Model 1, 2 and 3 because the inclination is closer to the critical value $i_{\rm c+}$ (Eq. \ref{eq:ic+}), while, in Model 5, the $\varpi$ precession rate is negative and its direction is retrograde.
The equivalent circular radius is very close to Saturn's radius in Model 4 ($a_{\rm eq}\sim1.1R_0$) and is smaller than it in Model 5 ($a_{\rm eq}\sim0.6R_0$).
Thus, we expect that a lot of material fall onto Saturn as a result of $e$ and $i$ damping in these cases.
The amount of surviving mass is expected to be much smaller than in the cases of Model 1, 2 and 3.
In this section, we demonstrate this predictions and provide a constraint on the $(i_{\rm TD}, q_{\rm TD})$ parameter space to avoid the significant loss of captured fragments by falling onto the planet.

\begin{figure}
\centering
\includegraphics[width=\linewidth]{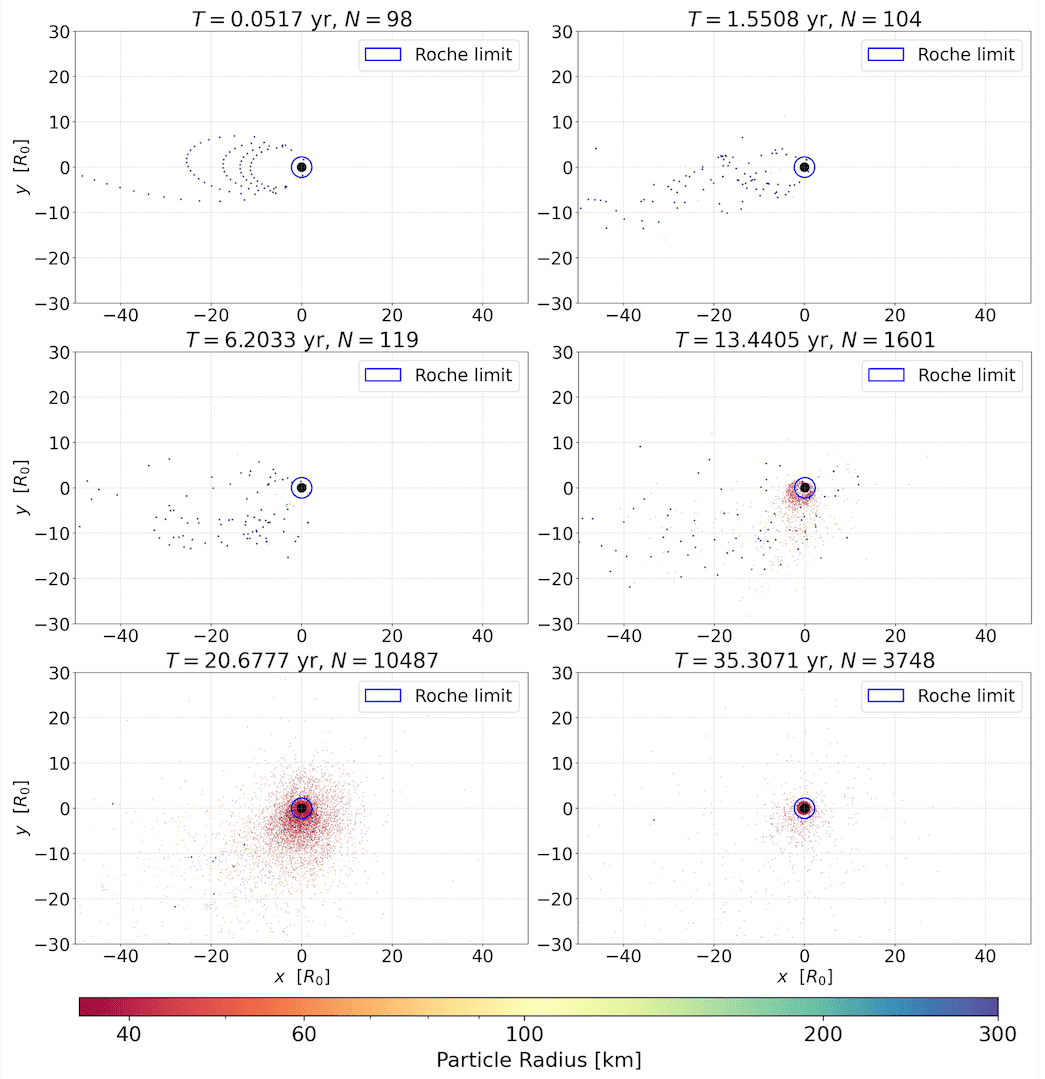}
\caption{Same as Fig. \ref{fig:snap_i1.0}, but for $i_{\rm TD}=45^{\circ}$ and $q_{\rm TD}=1.1 R_0$ (Model 4).}
\label{fig:snap_i45}
\end{figure}

\begin{figure}
\centering
\includegraphics[width=\linewidth]{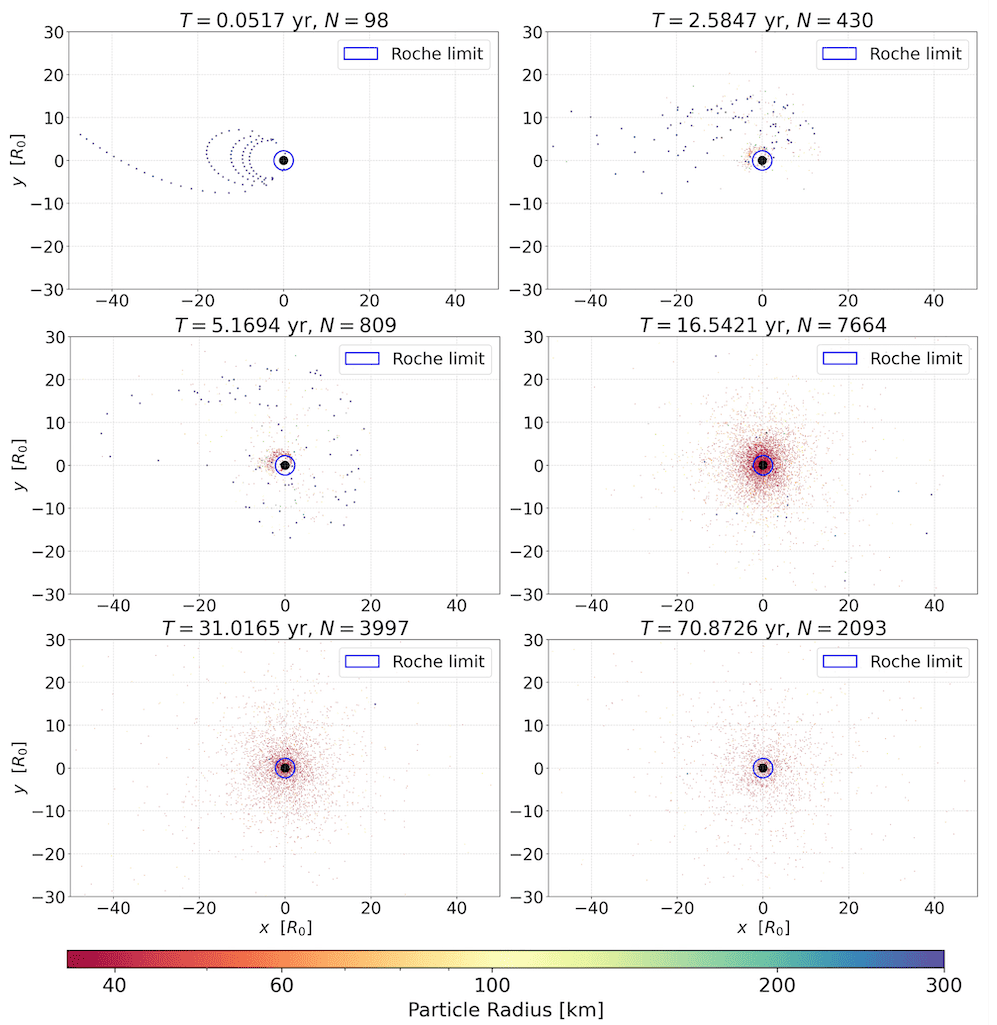}
\caption{Same as Fig. \ref{fig:snap_i1.0}, but for $i_{\rm TD}=60^{\circ}$ and $q_{\rm TD}=1.1 R_0$ (Model 5)}
\label{fig:snap_i60}
\end{figure}

Figure \ref{fig:snap_i45} and \ref{fig:snap_i60} show the snapshots of the time evolution in Model 4 and 5 (see also Fig. \ref{fig:snap_i45_zoom} and \ref{fig:snap_i60_zoom} for their zoom-up and Fig. \ref{fig:snap_i45_zoom_z} and \ref{fig:snap_i60_zoom_z} for these side view).
In Model 4, we find that the significant collisional grinding occurred before forming a torus-like structure by differential $\varpi$ precession, which is in contrast to the case of Model 2 where the torus-like structure firstly form following the collisional evolution (see Fig. \ref{fig:snap_i20}).
This is because the time to trigger the collisional grinding is shorter than the $\varpi$ precession timescale ($\sim70$ years) in spite of the larger spatial spread of particles.
On the other hand, in Model 5, the $\varpi$ precession direction is retrograde sense because of negative $\tau_{\varpi}$ (see Fig. \ref{fig:Precession Rate}).
After the onset of collisional grinding, the eccentricity and inclination of particles are damped by dissipative collisions. 
However, because the equivalent radius in these cases are very close to or smaller than Saturn's radius, a large amount of particles fall onto the planet and the number of particles decreases in our simulations.
Seen from Fig. \ref{fig:snap_i45_zoom} and \ref{fig:snap_i60_zoom}, much less massive ring with orbital radius close to the Saturn's radius forms in Model 4, while no ring-like structure forms in Model 5.

\begin{figure}
\centering
\includegraphics[width=0.8\linewidth]{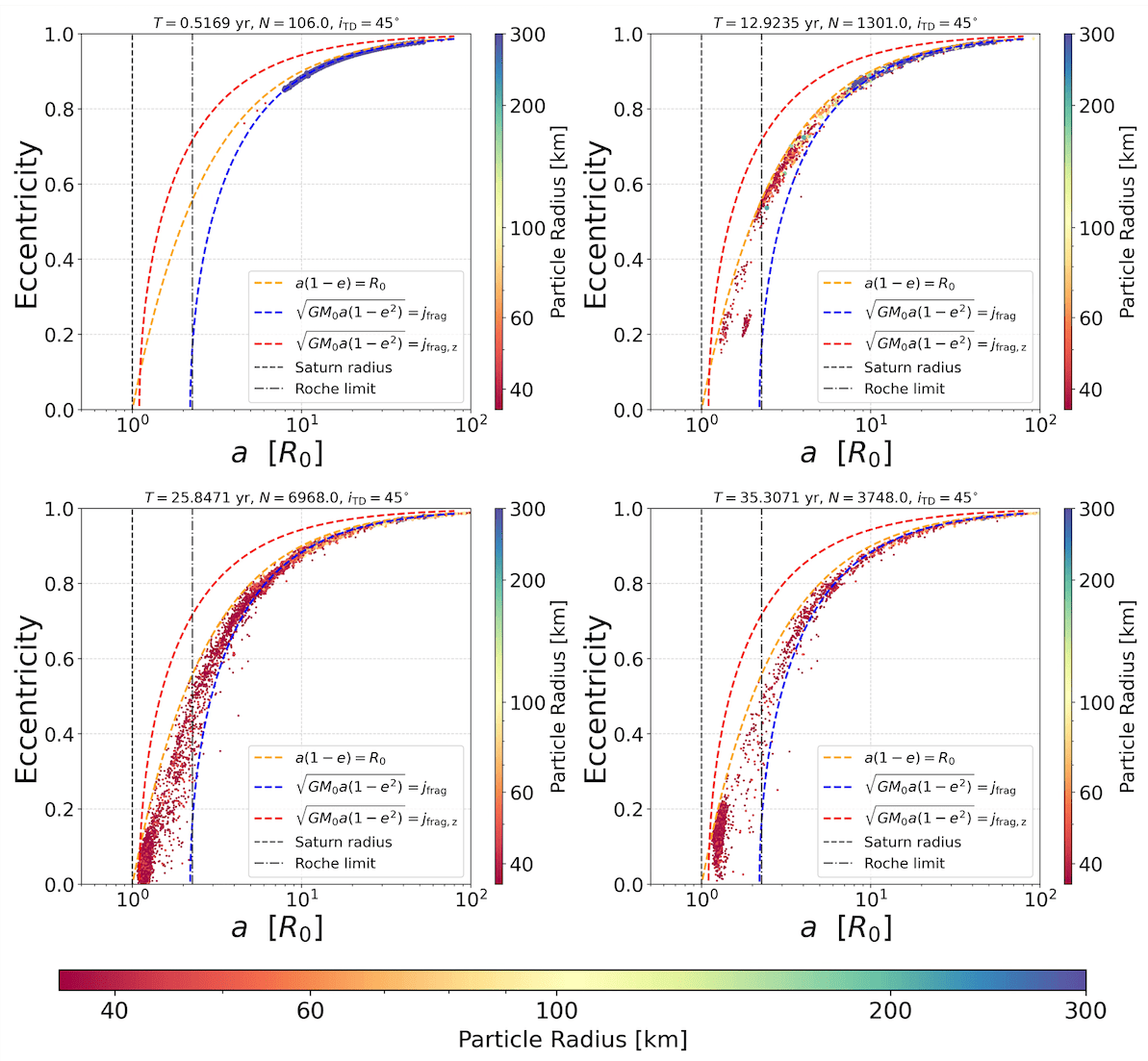}
\caption{The time evolution of $e-a$ distribution in Model 4. The blue and red curves represent the orbit with the angular momentum equal to the magnitude of angular momentum and $z$ component of angular momentum of the tidally disrupted body, respectively. The orange curve shows the orbit whose the pericenter distance is equal to the Saturn's radius. The dashed and dash-dotted black lines shows the Saturn's radius and its Roche limit. }
\label{fig:snap_ea_i45}
\end{figure}

Figure \ref{fig:snap_ea_i45} shows the time evolution of the $a-e$ distribution in Model 4.
The color of each particle represents their particle size.
The red and blue dashed curves correspond to $j(a, e)=j_{\rm frag, z}$ and  $j(a, e)=j_{\rm frag}$, respectively, where $j(a, e)=\sqrt{GM_0a(1-e^2)}$ is the specific angular momentum of particles and $j_{\rm frag, z}$ is the $z$ component of $j_{\rm frag}$.
Thus, the particles strictly lie on the blue dashed curve in the initial condition as described in Section \ref{sec:settings}.
Because of the conservation of $z$ component of the angular momentum, the particles deviate from the blue dashed curve and approach the red dashed curve as damping their eccentricity and inclination, which is consistent with the analytical argument in Section \ref{sec:EqRadius}.
During the evolution, the particles on the orbits whose pericenter distance is smaller than the planet radius are removed, thus the particles on $a-e$ space are confined by the orange dotted curve which represent the orbit with pericenter distance equal to the planet radius. 
As a result, the surviving mass decreases after the onset of collisional evolution.

\begin{figure}
\centering
\includegraphics[width=\linewidth]{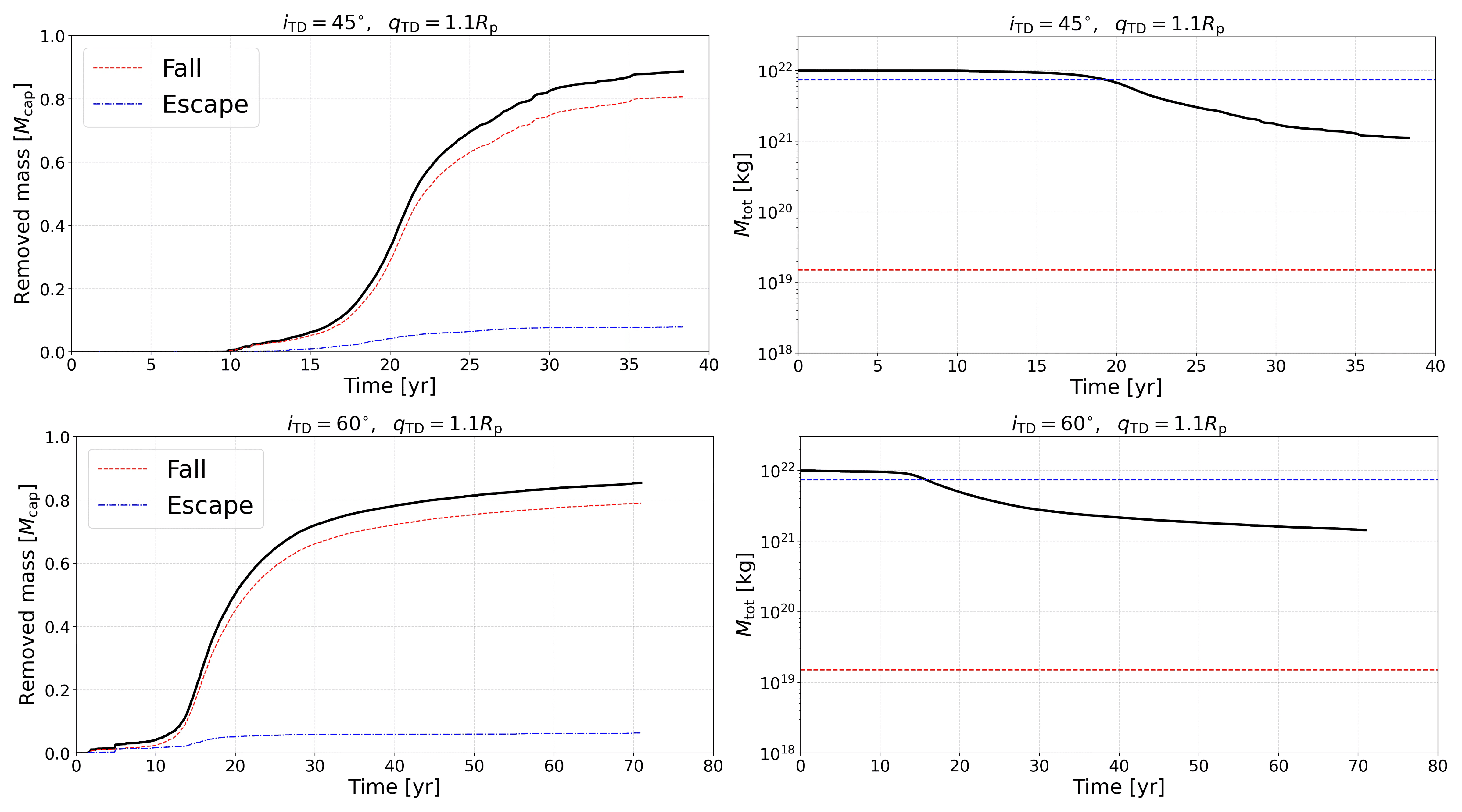}
\caption{The upper and lower left panels show the time evolution of removed mass in Model 4 and 5, respectively. The red dotted curve shows the particles falling onto the Saturn's surface and the blue dashed curve shows the particles escaping from the Saturn's gravity.
The upper and lower right panels show the time evolution of the surviving mass in Model 4 and 5, respectively. The red and blue dotted lines are same as that of Figure \ref{fig:eccDapming_Model1,2} and \ref{fig:incDapming_Model2}. }
\label{fig:RemovedMass}
\end{figure}

Figure \ref{fig:RemovedMass} shows the time evolution of the mass loss (the left panels) and surviving mass (the right panels) in Model 4 and 5.
The red dashed and blue dash-dotted curves on the left panels correspond to the removed particles which fall onto Saturn and the particles which reach the outer boundary ($r=1000 R_0$) in our simulation.
The number of removed particles rapidly increases just after the onset of collisional grinding ($\sim 10$ yr).
We find that $\gtrsim80\%$ of the captured fragments are removed during the $e$ and $i$ damping in both models, most of which are removed because of falling onto Saturn.
The blue and red dotted lines in the right panels represent the mass including the present Saturn's rings and inner satellites and the mass of the present Saturn's rings, respectively.
The right panels show that enough mass to form the present ring-satellites system of Saturn by mechanism proposed by  \cite{Crida:2012} cannot be supplied in these cases
\footnote{
We note that the mass of the surviving fragments depends on the mass of the initially tidally-disrupted object.
In our simulation, the mass of the disrupted body is assumed to be $\sim 10$ Pluto mass, which is close to the upper end of size distribution of primordial KBOs \citep{Nesvorny:2016}. 
If we consider a smaller disrupted body, the surviving mass becomes smaller.
Thus, our estimate of the surviving mass is considered to be an upper limit for a single event.
}.
Therefore, we conclude that, in order to form a circular and equatorial ring-like structure with enough mass including present rings and mid-sized satellite by a single tidal disruption event, its inclination $i_{\rm TD}$ and pericenter distance $q_{\rm TD}$ are required to be small enough (roughly saying, $i_{\rm TD}\lesssim 45^{\circ}$ and $R_0\lesssim q_{\rm TD}\lesssim 1.4R_0$ in the case of Saturn, see also Fig. \ref{fig:qi_space_constraint}).

\subsection{Classifying the dynamical path and possibility of the ring formation on the $(q_{\rm TD}, i_{\rm TD})$ parameter space} \label{sec:constraint}

\begin{figure}
\centering
\includegraphics[width=0.8\linewidth]{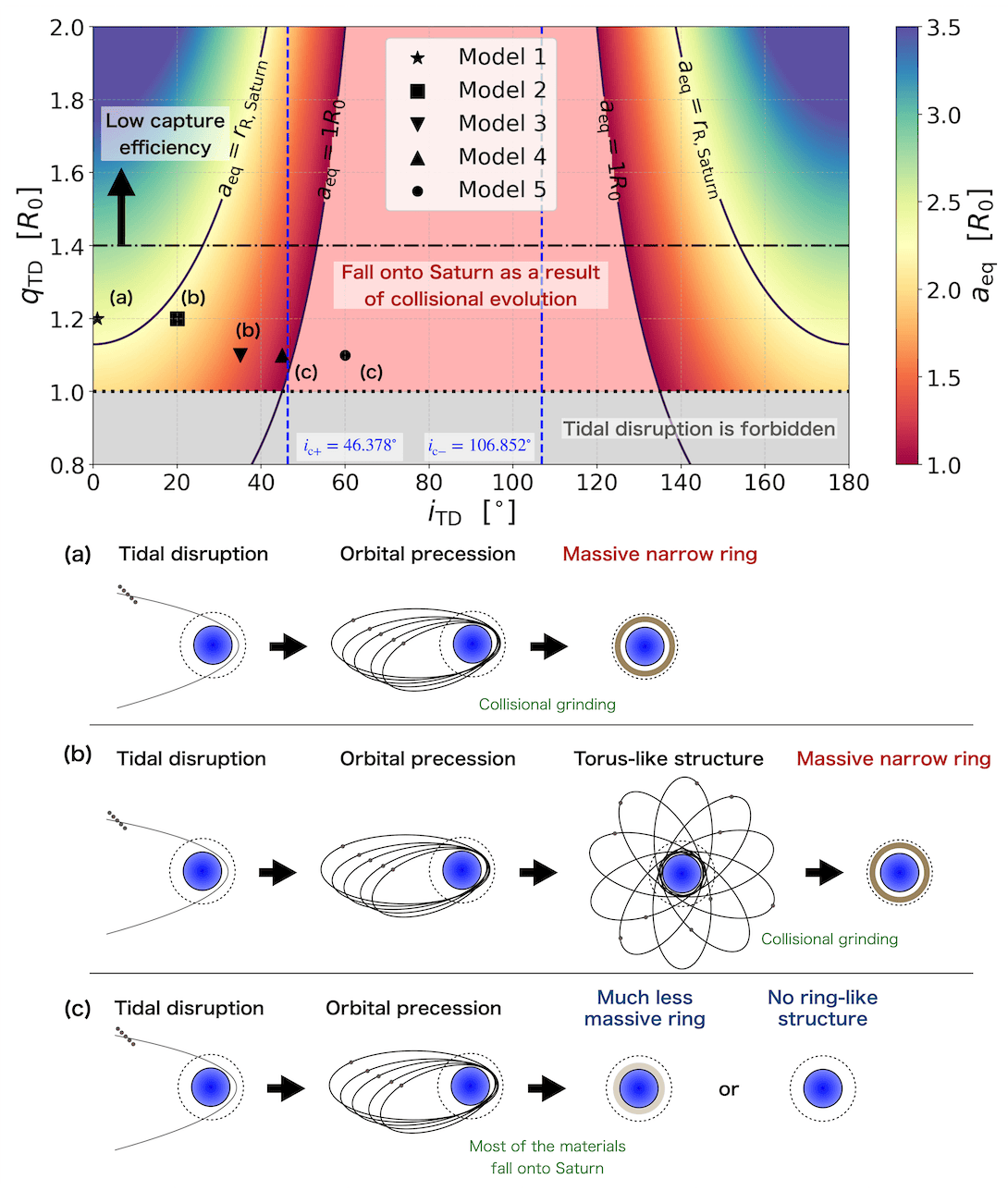}
\caption{The summary of classification of the dynamical path and fate of the tidal fragments.
The upper panel shows $(i_{\rm TD}, q_{\rm TD})$ parameter space colored with the equivalent radius (Eq. \ref{eq:a_eq}).
The grey shaded region represents where the pericenter distance is smaller than Saturn's radius and tidal disruption is forbidden.
The red shaded region represents where the equivalent orbital radius is smaller than Saturn's radius and significant amount of fragments fall onto Saturn as a result of collisional evolution. 
The upper region from black dash-dotted line represents where the capture efficiency of tidal disruption is less than $\sim1\%$ from SPH simulations by \cite{Hyodo:2017}. 
The parameter values of each model are also plotted with different markers.
The numbers close to the marker correspond to the classified dynamical path in our simulations schematically depicted in the lower part of this figure.
The blue dashed line shows the critical inclination $i_{c+}$ and $i_{c-}$ (Eq. \ref{eq:ic+} and \ref{eq:ic-}).}
\label{fig:qi_space_constraint}
\end{figure}

Compiling the results of $N$-body simulation results in the previous sections and SPH simulations by \cite{Hyodo:2017}, 
We classify the dynamical path and fate of captured tidal fragments after the tidal disruption on the $(q_{\rm TD}, i_{\rm TD})$ parameter space.
Figure \ref{fig:qi_space_constraint} illustrates the classification on the parameter space and the dynamical path is schematically depicted in the lower part of the figure.
The upper panel represents the equivalent circular radius calculated with Eq. \eqref{eq:a_eq} by color map.
The closer the inclination of tidal disruption is to $90^{\circ}$, the smaller the equivalent orbital radius becomes, because of the conservation of $z$ component of the angular momentum.
The red shaded region represents where the equivalent orbital radius (Eq. \ref{eq:a_eq}) is smaller than Saturn's radius thus a large amount of captured fragments are expected to fall onto Saturn as a result of collisional damping of their eccentricity and inclination.
The grey shaded region represents where the tidal disruption is prohibited because the pericenter distance is smaller than Saturn's radius, that is, a direct collision to Saturn occurs rather than tidal disruption.
In the upper region of the black dash-dotted line, the capture efficiency by the tidal disruption becomes smaller than 1\% in SPH simulation by \cite{Hyodo:2017}.
The $(i_{\rm TD}, q_{\rm TD})$ values in each model are also plotted in Fig. \ref{fig:qi_space_constraint} with different markers.
The dynamical path of each model is represented by the number in the vicinity of each marker, which corresponds to the schematic illustration of the dynamical path (a), (b) and (c) in the lower part of this figure.
The blue dashed line shows the critical inclinations $i_{\rm c+}$ and $i_{\rm c-}$ (Eq. \ref{eq:ic+} and \ref{eq:ic-}) at which the $\varpi$ precession timescale becomes infinity.

For the prograde cases ($i_{\rm TD}\lesssim 90^{\circ}$), when the inclination is small (Model 1), the vertical dispersion of the tidal fragments is small and the collision probability among them is large.
Therefore, the collisional grinding is triggered before the $\varpi$ precession is completed and circlarized rings can form before forming torus-like structure (path (a)).
When the inclination is larger but not close to $i_{\rm c}$ and the inclination with $a_{\rm aq}=1R_0$ (Model 2 and 3), a torus-like structure firstly forms following the onset of collisional grinding (path (b)), which is consistent picture with \cite{Hyodo:2017}.
The collisional evolution proceeds with conserving $z$ component of the system's angular momentum.
As a result, the formed ring has a radially narrow structure and its orbital radius is determined by $(q_{\rm TD}, i_{\rm TD})$ (see Section \ref{sec:NarrowRings}).
Most of the tidal fragments can survive the collisional evolution in both of path (a) and (b) because the equivalent radius is much larger than the Saturn's radius, thus enough mass to form the current ring and inner satellites of Saturn as proposed by \cite{Crida:2012} can be supplied even by only one tidal disruption event of a 10 Pluto mass body (see Section \ref{sec:DynamicalPath}).
When the inclination is close to the inclination with $a_{\rm aq}=1R_0$ (Model 4 and 5), most of the captured fragments fall onto Saturn (path (c)).
In particular, when the equivalent radius is smaller than the Saturn's radius, no ring-like structure can form.
Thus, massive rings cannot form only by a single event in these cases (see Section \ref{sec:larger_i}).

Our $N$-body simulations presented here is only for the prograde case.
For the retrograde case, $a_{\rm eq}$ is symmetric with respective with $i_{\rm TD}=90^{\circ}$.
When $90^{\circ}< i_{\rm TD}< 135^{\circ}$, almost all fragments fall onto Saturn and ring structure does not form.
When $135^{\circ}< i_{\rm TD}< 180^{\circ}$, the $\varpi$ precession rate is about twice larger than the prograde case (see Fig. \ref{fig:Precession Rate}).
Thus, a torus-like structure forms before the onset of collision evolution.

These classifications of dynamical path and orbital radius shown in Fig. \ref{fig:qi_space_constraint} are consistent with our direct $N$-body simulations present in Section \ref{sec:DynamicalPath} to \ref{sec:larger_i}.
Thus, the inclination $i_{\rm TD}$ and pericenter distance $q_{\rm TD}$ are the key parameters to the dynamical evolution followed by the tidal disruption.

\section{Discussion}\label{sec:Discussion}
\subsection{Implication on the primordial ring-satellite system evolution}\label{sec:Implication_narrow}
In Section \ref{sec:NarrowRings}, we showed that the finally-circularized ring is radially narrow at the orbital radius $\sim a_{\rm eq}$.
This interesting results may bring us an important implication on the initial condition of primordial ring-satellite system evolution, thus it is a key to connecting between ring and satellite formation.
\cite{Salmon:2010} performed a long-term viscous evolution of dense rings around Saturn with 1D viscous evolution model.
Their initial condition was set to be a narrow ring with a gaussian surface density profile, which is very similar to the finally formed rings in our simulations.
They showed that, when the ring is well inside the Roche limit, the rings viscously spread and the narrow gaussian profile is flattened.
On the other hand, when the ring is close to the Roche limit, the outer edge of spreading ring reached the Roche limit faster and more amount of mass are removed from the outer edge.
They suggested that the removed mass from the Roche limit is accumulated and transformed into aggregates which gravitationally interacts with the remaining rings.
However, their simulation did not include such an effect because they focused only on the evolution of the ring itself.
So far, several studies have investigated the coupled evolution of spreading rings with the newly formed satellites \citep{Charnoz:2010, Charnoz:2011, Crida:2012, Salmon:2017}.
For example, \cite{Charnoz:2010} performed a hybrid simulation self-consistently coupling a 1D hydrodynamical model for the spreading rings with physically realistic ring viscosity \citep{Daisaka2001} and analytical orbital model for the formed aggregates.
As a results, they succeeded in quantitatively reproducing the trend that the small moon's mass orbiting just outside the main ring increases with its orbital radius.
\cite{Crida:2012} extended this idea, and proposed that the regular satellites around Saturn, Neptune and Uranus were formed from ancient massive rings in a similar way.
Their model was also able to excellently reproduce the relation between satellite mass and its orbital radius (see Fig 1. in \cite{Crida:2012}).
However, the initial rings are assumed to have a constant surface density extending from close to the planet surface up to its Roche limit in their simulations.

Our analytical arguments described in Section \ref{sec:EqRadius} combining our $N$-body simulations indicate that the initial surface density profile is neither broad nor flat but a narrow structure with its radius determined by the inclination $i_{\rm TD}$ and pericenter distance $q_{\rm TD}$ of the tidal disruption.
While there exist a parameter space where the narrow ring centered at around the middle of the region within the Roche limit is created, it is also possible to form the narrow ring centered at beyond the Roche limit (see Fig. \ref{fig:qi-space}).
We expect that, in the latter case, clumps would be directly formed from the material located near or outside the Roche limit radius and finally satellites can form, while, in the former case, the narrow ring spreads by ring viscosity and leads to the satellite formation \citep{Crida:2012}.
In the latter case, if most of mass are accumulated and transformed into the satellites, the tidal disruption may be also considered as a satellite formation process together with the ring formation process \citep{Kegerreis:2025}.
Although the detail investigation of the viscous long-term evolution including the accumulation of satellites from the narrow ring is beyond the scope of this paper, this topic seems to be worth being studied in more detail.

\subsection{The collision velocity}\label{sec:CollisionVel}
\begin{figure}
\centering
\includegraphics[width=\linewidth]{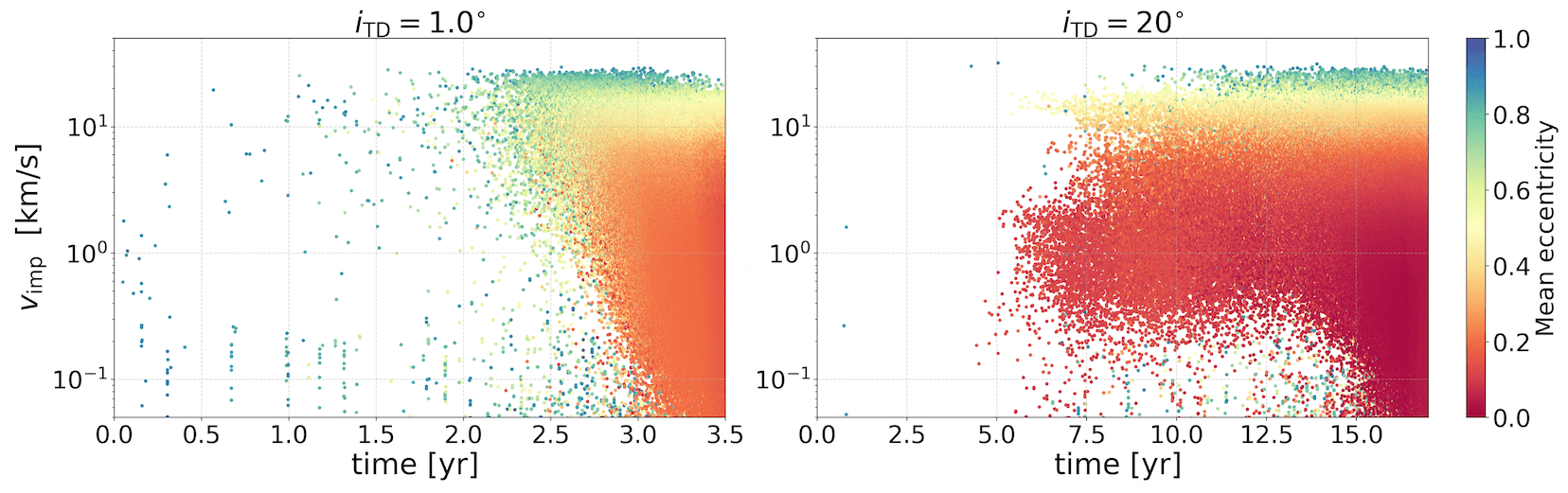}
\caption{The collision velocity occurring in Model 1 ($i_{\rm TD}=1.0^{\circ}$) and Model 2 ($i_{\rm TD}=20^{\circ}$). The color of each plot represents the mean eccentricity of impactor and target particles.}
\label{fig:vimp}
\end{figure}

Here, we discuss the collision velocity between fragment particles during dynamical precession.
The left and right panels in Figure \ref{fig:vimp} show the collision velocity ($v_{\rm col}$) distribution with time in Model 1 and 2.
The color of each plot represents the mean eccentricity defined as $e_{\rm mean}=(e_{\rm imp}+e_{\rm tar})/2$, where $e_{\rm imp}$ and $e_{\rm tar}$ are the eccentricity of the impactor and target particles, respectively.

Because, in Model 1 ($i_{\rm TD}=1.0^{\circ}$), the system is almost on the equatorial plane and the onset time of the collisional grinding is so early that $\varpi$ is not fully randomized and almost aligned, the collision velocity at the initial phase ($\lesssim 1$ yr) is relatively small ($\sim 1 $ km/s) despite its high mean eccentricity.
After the alignment of $\varpi$ is broken, the high-velocity collisions ($\gtrsim 10$ km/s) can occur between particles with high eccentricity.
Thus, the increase trend of the collision velocity at initial phase ($\lesssim 1.0$ yr) is found in the left panel of Figure \ref{fig:vimp}.
After that, the damping of particles eccentricity continues to proceed, as a result, low-velocity collisions ($\sim 0.1-1$ km/s) can occur between particles with small eccentricity at later phase ($\gtrsim 2.5$ yr). 
On the other hand, in Model 2 ($i_{\rm TD}=20^{\circ}$), the high-velocity collisions could occur even at the moment of the onset of collisional grinding when $\varpi$ is already fully randomized.
Because the high-velocity collision strongly dissipates the energy and damp the eccentricity significantly, many low-velocity collisions ($\sim 1$ km/s) with small mean eccentricity occur just after a first few high-velocity collisions.

\begin{figure}
\centering
\includegraphics[width=\linewidth]{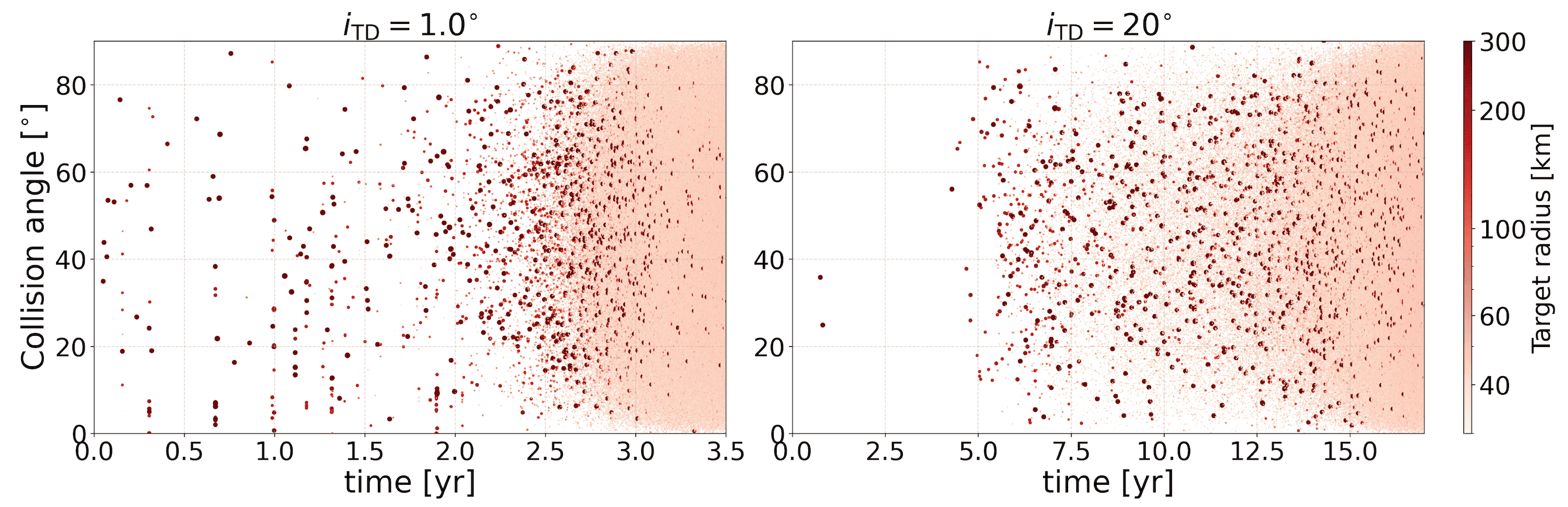}
\caption{The collision angle distribution in Model 1 ($i_{\rm TD}=1.0^{\circ}$) and Model 2 ($i_{\rm TD}=20^{\circ}$). The color and size of each plot represents the target radius.}
\label{fig:colAngle}
\end{figure}

\begin{figure}
\centering
\includegraphics[width=\linewidth]{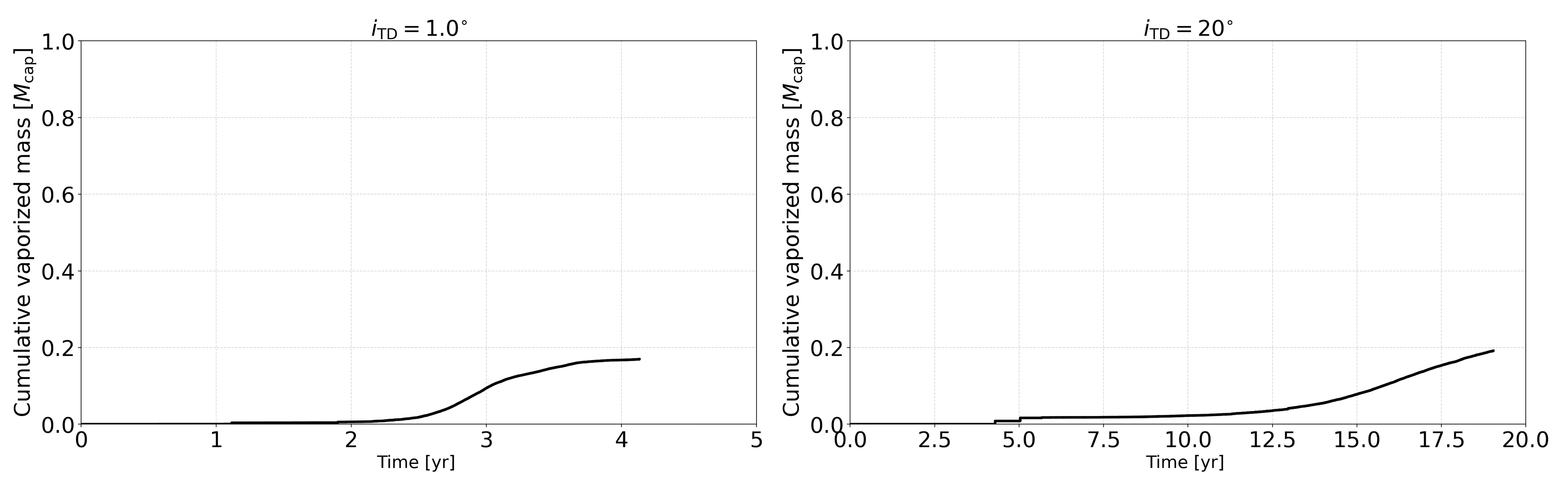}
\caption{The cumulative mass of impactor particles which have experienced the high-velocity collisions with shock pressure larger than 70 GPa in Model 1 (left panel) and Model 2 (right panel). Here, we assume that the mass of $M_{\rm t}\sin \theta_{\rm col}$ is completely vaporized when $P_{\rm shock}>P_{\rm vap}$.}
\label{fig:Pshock_cumulative}
\end{figure}

While in our simulations, the phase change of fragments during a collision such as melting or vaporization has not been directly incorporated, we can discuss it by converting the kinetic energy of collisions to the thermal energy during the collision.
We can evaluate its importance by comparing the shock-induced pressure $P_{\rm shock}$ within the colliding material during the collision event in the simulation with the critical shock pressure $P_{\rm vap}$ for the phase change shown in Table 1 of \cite{Kraus:2011} based on the experiment by \cite{Stewart:2008}.
The former is evaluated by $v_{\rm col}$ obtained in the simulation as \citep[e.g.,][]{Hyodo:2021}
\begin{equation}
    P_{\rm shock}=\rho_0(C_0+su_{\rm p})u_{\rm p},\label{eq:Pshock}
\end{equation}
where $\rho_0$ is the material density before the shock compression, $C_0=3.0$ km/s is the bulk sound velocity for icy material, $u_{\rm p}\simeq0.5 v_{\rm col}$ is the particle velocity of the shocked material, $s=\dd{u_{\rm s}}/\dd{u_{\rm p}}\simeq 1$ is a constant and $u_{\rm s}$ is the shock velocity.
According to Table 1 in \cite{Kraus:2011}, $P_{\rm vap}$ for complete vaporization is $\sim 70$ GPa using 5-phase equation of state \citep{Senft:2008}.
To roughly estimate how large amount of material has experienced the vaporization induced by a collision, we adopt $P_{\rm vap} = 70 \, \rm GPa$ as a threshold value and plot the cumulative impactor mass which has experienced the complete vaporization in Figure \ref{fig:Pshock_cumulative}.
Here, considering the effect of the collision angle (Fig. \ref{fig:colAngle}), we assume that the mass of $M_{\rm t}\sin \theta_{\rm col}$ is completely vaporized when $P_{\rm shock}>P_{\rm vap}$.
From Fig. \ref{fig:Pshock_cumulative}, in Model 1 and 2, only less than $20 \%$ of the captured material ($\lesssim 2\times10^{21}$ kg) has experienced the complete vaporization within the simulated timescale.
Thus, we expect that the effect of vaporization on the dynamical evolution would not be dominant and it may not change our conclusion.
Future works should study the effects which are not considered here, such as vapor production, recondensation and other complex physics in vaporization.

Here, we discuss the expected evolution of the vaporized material, although the detail study is left for future works.
A high-velocity collision inducing the vaporization generates a vapor cloud, then it expands and cools adiabatically.
As a result, the vaporized material may be quickly recondensed.
According to the results of \cite{Hyodo:2025a}, the size of condensates becomes much smaller than those of colliding bodies.
The orbit of such a small condensate particle is vulnerable to non-gravitational forces such as the solar radiation pressure and Poynting-Robertson force, as a result, they could be removed by falling onto the central planet \citep[e.g., ][]{Liang:2023}.
Therefore, the actual survival fraction of the captured material discussed in Section \ref{sec:larger_i} could be reduced further.
However, the simulation including the effects of vaporization followed by recondensation process and non-gravitational dynamics of the condensates is very complicated 
and beyond the scope of this paper.

\subsection{The expected planetocentric orbital inclination of a passing body}
\label{sec:Inclination}
In the previous sections, through analytical calculation and $N$-body simulations, we derived the region on $(q_{\rm TD}, i_{\rm TD})$ parameter space where enough tidal debris can be captured around the planet and it can survive the subsequent collisional evolution.
From the conservation of the $z$ component of angular momentum around the planet, we showed that small inclination of the passing body (at least, $i_{\rm TD}\lesssim 45^{\circ}$) is required to supply enough material to form the Saturnian ring-satellite system by tidal disruption (see Fig. \ref{fig:qi-space}).
Here, we roughly discuss the expected distribution of $i_{\rm TD}$ from results of previous simulations.

\subsubsection{Giant planets}\label{sec:GiantPlanet}
\cite{Nesvorny:2007} investigated the capture of irregular satellites around giant planets during close encounters.
They investigated each close encounter with giant planets during LHB phase, based on the results of original ``Nice model'' simulation by \cite{Gomes:2005}
\footnote{
They obtained the planetocentric orbital elements of the dynamically stable irregular satellites around Jupiter, Saturn, Uranus and Neptune (see their Figures 4-7).
However, the inclination distribution of the dynamically stable irregular satellites in \cite{Nesvorny:2007} may not directly reflect that of initially injected body into the planet’s Hill sphere.
This is because the irregular satellites with inclination close to $90^{\circ}$ are dynamically unstable due to Kozai resonance \citep[see also][]{Nesvorny:2003}, resulting in its ejection of an irregular satellite or its collision with the central planet.}.
They suggested that the inclination distribution is proportional to $\sin i$ once the planetesimal disk becomes dynamically excited during planetary encounters \citep{Nesvorny:2007}.
We therefore expect tidal disruptions with $i<45^{\circ}$ (prograde) or $i>135^{\circ}$ (retrograde) to occur sufficiently frequently that ring formation may have been possible during the LHB era. 
Of course, the inclination distribution could depend on several factors such as the migration model of giant planets triggering LHB and planetesimal disk parameters, which is left for future work.

As discussed above, a close encounter can also occur on a retrograde planetocentric orbit ($i>90^{\circ}$). 
In such a case, retrograde tidal fragments may be captured, potentially leading to the formation of a retrograde ring. 
However, all known rings around the giant planets in the Solar System are prograde. 
Therefore, if tidal disruption is responsible for the formation of all giant-planet rings, this would imply either that only prograde encounters occurred by chance, or that some physical mechanism preferentially removes retrograde rings, leaving only prograde rings at the present day. 
This issue is beyond the scope of the present study.
In addition, as calculated by \cite{Hyodo:2017}, giant planets may have had chance to experience close encounter with KBOs at least a few times in the LHB era.
If a retrograde tidal disruption occurs with already formed prograde rings, most of material may fall onto the planet as a result of the collisional interaction between retrograde and prograde fragments, because their directions of $z$ component of the angular momentum vector are opposite and they are partially canceled out through collisions.
The collisional interaction between prograde and retrograde fragments after the tidal disruptions may be worth studying, which is left for future works.

\subsubsection{Terrestrial planets}
Recently, \cite{Kegerreis:2025} proposed a new formation scenario of Mars's small moons, Phobos and Deimos.
In their scenario, Mars captured some fraction of tidal fragments through the tidal disruption of a passing asteroid and the subsequent collisional evolution as discussed in this paper resulted in the formation of debris disk originating Phobos and Deimos.
They modeled the tidal disruption with SPH simulations and the subsequent long-term orbital evolution with $N$-body simulations without collisional fragments as studied by \cite{Hyodo:2017}.
Through a series of simulations, they showed that a few percent of the parent asteroid contributes to the formation of debris disk after the collisional evolution, thus it is possible to supply the enough material to form Phobos and Deimos with an appropriate range of parent asteroid's mass.
They did not consider the effect of inclination on the collisional evolution as discussed here, but our results indicate that the orbital radius of the debris disk significantly depends on it.
According to the previous simulations of terrestrial planet formation, the impact angle distribution in giant impact phase was expected to be almost isotropic \citep[e.g.,][]{Agnor:1999, Chambers:2001, Kokubo:2007}.
Thus, if we assume that the Mars' moon formation event by tidal disruption occurred in the giant impact phase, we argue that a debris disk with a wide range of orbital radius from Mar's surface to or beyond its Roche limit can form depending on the pericenter distance and inclination, which may affect the final orbital structure of the satellite system.

Interestingly, \cite{Tomkins:2024} recently suggested that the Earth had a ring system in the past as well.
They investigated the 21 asteroid impact craters on Earth recorded in the period with a dramatic increase in impact rate in the mid-Ordovician  \citep[$\sim 480$ Ma, e.g.,][]{Schmitz:2001}, and found that all of the craters were located in an equatorial region at $\leq 30^{\circ}$ in paleolatitude.
They proposed that this feature was produced by the falling material from the equatorial debris ring around the Earth originated from tidal disruption.
They did not perform any dynamical simulation of tidal disruption and collisional evolution around the Earth, but our simulations can also be applied to this problem.

In summary, it is worth considering the dynamical fate of tidal debris around the terrestrial planets as well, despite the smaller Hill radius of the terrestrial planets than that of giants planets due to the lower mass and closer location to the Sun.
Our simulation can directly apply to the terrestrial planets as well as to the other celestial objects.

\section{Summary and Conclusion}\label{sec:Summary}
While the ring system is common in giant planets and small celestial bodies in our Solar System, most of the proposed formation scenarios focused on the Saturn's rings.
The tidal disruption scenario is applicable to the other giant planets \citep{Dones:1991, Hyodo:2017}.
In this scenario, a nearby passing body such as a KBO enters deep into the Roche limit of the giant planet and is tidally disrupted by its strong tidal force.
Some fraction of the fragments are gravitationally bound by the planet in a highly-eccentric planetocentric orbit. 
The pericenter and ascending node of these orbits precess due to the $J_2$ term of the planet's gravity potential.
As a result, the collisional grinding is triggered by the orbital crossings due to the differential precession, damping their eccentricity and inclination of the fragments and finally forming circular and equatorial rings.

\cite{Hyodo:2017} performed SPH simulations of the tidal disruption of a passing KBO nearby a giant planet and showed that 0.1-10 \% of the disrupted material is gravitationally captured around the planet.
They ended the simulations just after the disruption and discussed the following dynamical evolution of the captured fragments with analytical arguments and non-collisional $N$-body simulations.

In this paper, we performed the $N$-body simulations of the dynamical evolution of tidal fragments with collisional fragmentation.
We also revisited the analytical arguments on the precession rate and equivalent radius including the effects of inclination $i_{\rm TD}$ and pericenter distance $q_{\rm TD}$ of the orbit of an incoming body to be tidally disrupted.
Our main findings are as follows:
\begin{itemize}
    \item The analytical derivation shows that the precession rate significantly depends on $i_{\rm TD}$.
    In particular, the $\varpi$ precession timescale $\tau_{\varpi}$ becomes extremely long when the inclination is close to the critical value $i_{\rm c+}\sim46.378^{\circ}$ and $i_{\rm c-}\sim106.852^{\circ}$ (Fig. \ref{fig:Precession Rate}). 
    Because the $z$ component of the angular momentum is conserved, the equivalent radius $a_{\rm eq}$ after the circularization depends on $i_{\rm TD}$ and $q_{\rm TD}$, ranging from the planet radius to or beyond its Roche limit (Fig. \ref{fig:qi-space}). 
    \item The $N$-body simulation shows that the collisional grinding starts before forming the torus-like structure in the case of $i_{\rm TD}=1.0^{\circ}$ (Fig. \ref{fig:snap_i1.0}), while the torus-like structure first forms, followed by the significant collisional grinding in the case of $i_{\rm TD}=20^{\circ}$ and $35^{\circ}$  (Fig. \ref{fig:snap_i20} and \ref{fig:snap_i35}).
    Because the formation timescale of a torus-like structure by the differential precession is equivalent to $\tau_{\varpi}$, the collisional grinding starts before the formation of the torus-like structure when the inclination is close to $i_{\rm c}$ (Fig. \ref{fig:snap_i45}).
    The dynamical picture described by \cite{Hyodo:2017} is modified depending on the inclination $i_{\rm TD}$.
    \item The circularized equatorial rings are confined to a narrow radial region at the equivalent radius $a_{\rm eq}$ that is analytically calculated by Eq.~\eqref{eq:a_eq} (Fig.~\ref{fig:SurfaceDensity}).
    When $a_{\rm eq} \gtrsim R_0$, the narrow rings would diffuse in a longer timescale and lead to the satellite formation from the spreading rings \citep{Crida:2012}.
    On the other hand, when $a_{\rm eq} \lesssim R_0$,
    most of captured fragments are removed by falling onto the planet (Fig.~\ref{fig:RemovedMass}).
    Thus, to form a massive ring, $i_{\rm TD}$ of the passing body needs to be small enough that $a_{\rm eq}$ is well larger than $R_0$.
\end{itemize}

We compiled these findings on the $(q_{\rm TD}, i_{\rm TD})$ parameter space in Fig.~\ref{fig:qi_space_constraint}.
We found that $i_{\rm TD}$ and $q_{\rm TD}$ are the two key parameters to the dynamical evolution path and fate of the captured tidal fragments.
We stress that our analytical arguments and $N$-body simulations results are applied not only to Saturn but also to the other giant planets and terrestrial planets (see Section \ref{sec:Inclination}) as long as the treatment in the gravity regime of collision outcome model is valid.

\section{Acknowledgments}
We thank the two reviewers for their careful reading of our paper and insightful comments.
This work was supported by JSPS KAKENHI Grant Number JP25KJ1252.
R.H. acknowledges the financial support of JSPS Grants-in-Aid (23KK0253, 22K14091, 21H04512, 21H04514, 20KK0080, 26K00756).

\printcredits

\bibliographystyle{cas-model2-names}

\bibliography{reference}

\bio{}
\endbio

\endbio

\section*{Appendix A. Additional snapshots}\label{app:snap}
\begin{figure}
\centering
\includegraphics[width=\linewidth]{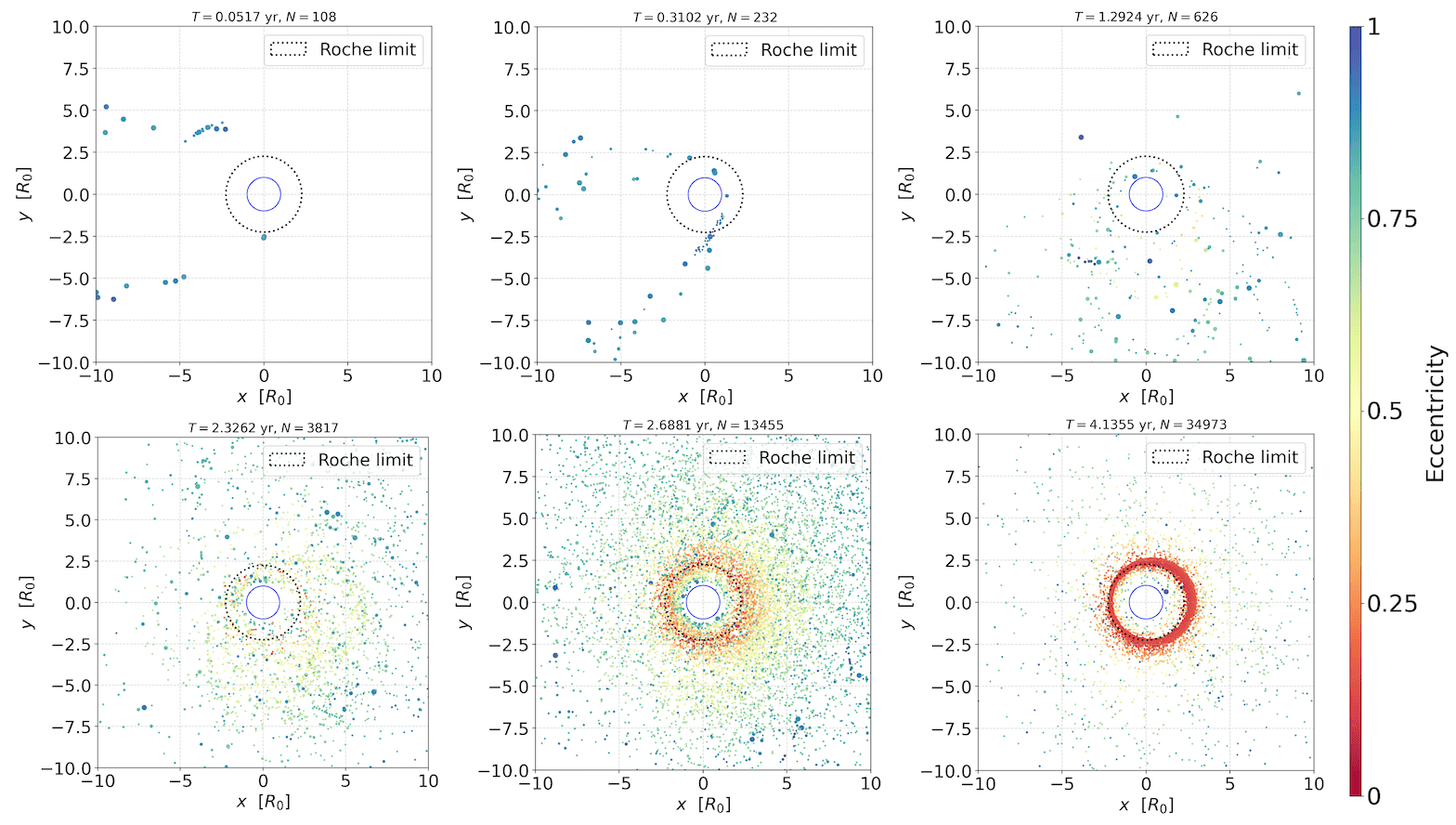}
\caption{The zoom-in view of Figure \ref{fig:snap_i1.0} (Model 1), but the color of each plot represents the particles eccentricity and their size reflect the particle radius.}
\label{fig:snap_i1.0_zoom}
\end{figure}

\begin{figure}
\centering
\includegraphics[width=\linewidth]{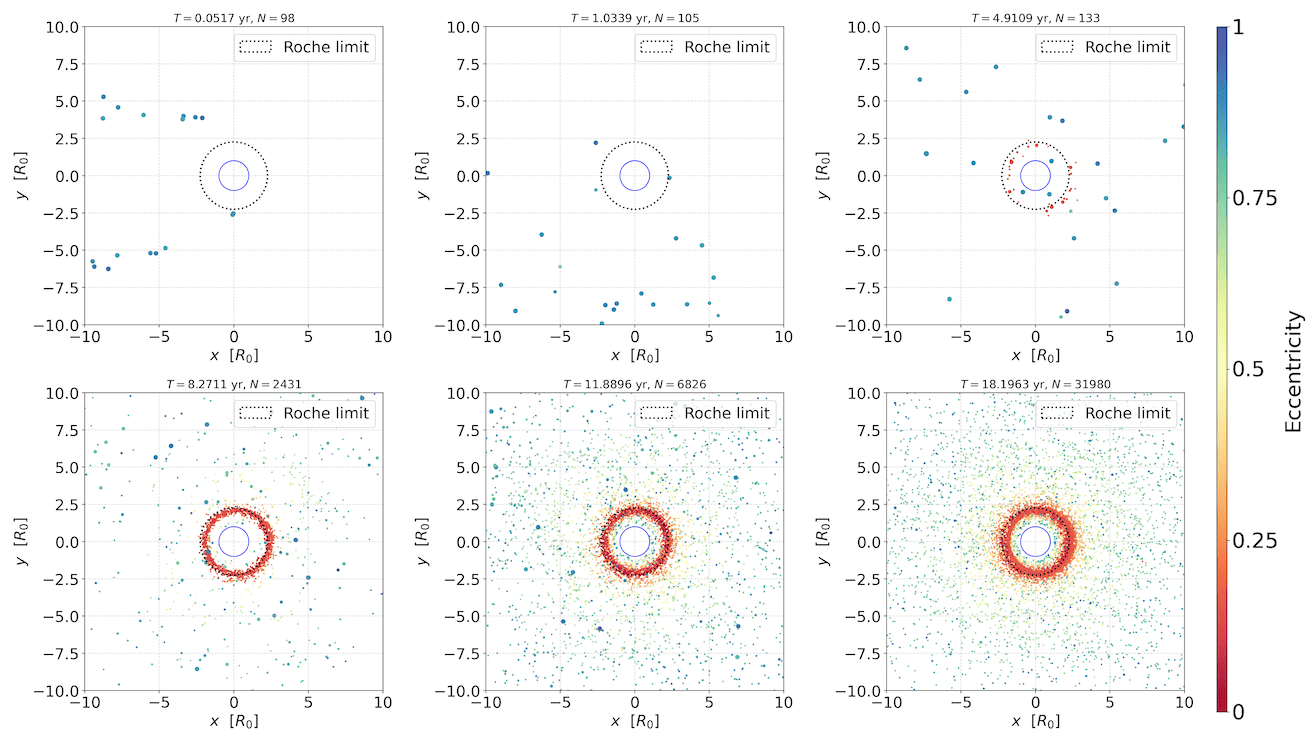}
\caption{The zoom-in view of Figure \ref{fig:snap_i20} (Model 2), but the color of each plot represents the particles eccentricity and their size reflect the particle radius.}
\label{fig:snap_i20_zoom}
\end{figure}

\begin{figure}
\centering
\includegraphics[width=\linewidth]{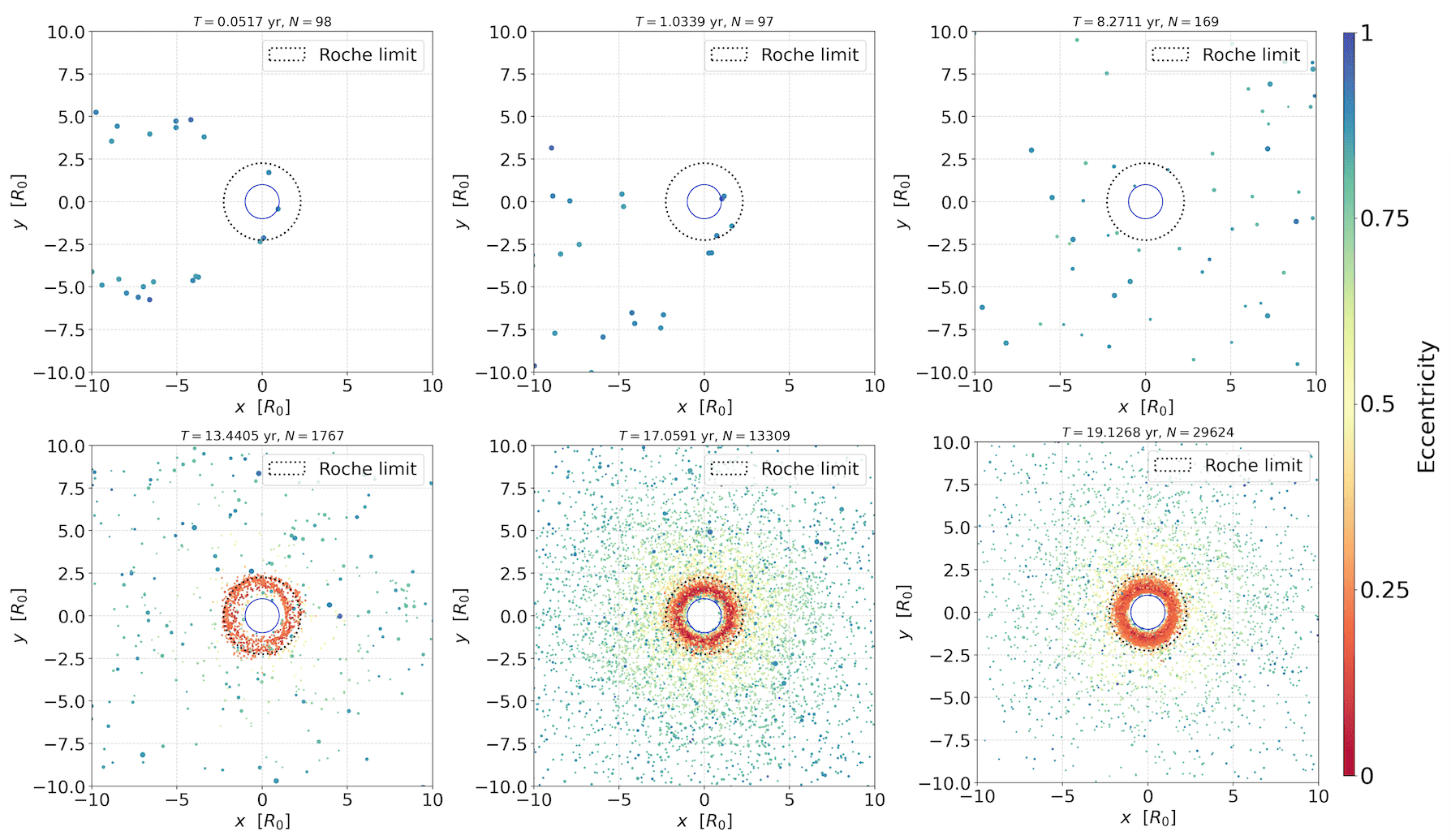}
\caption{The zoom-in view of Figure \ref{fig:snap_i35} (Model 3), but the color of each plot represents the particles eccentricity and their size reflect the particle radius.}
\label{fig:snap_i35_zoom}
\end{figure}

\begin{figure}
\centering
\includegraphics[width=\linewidth]{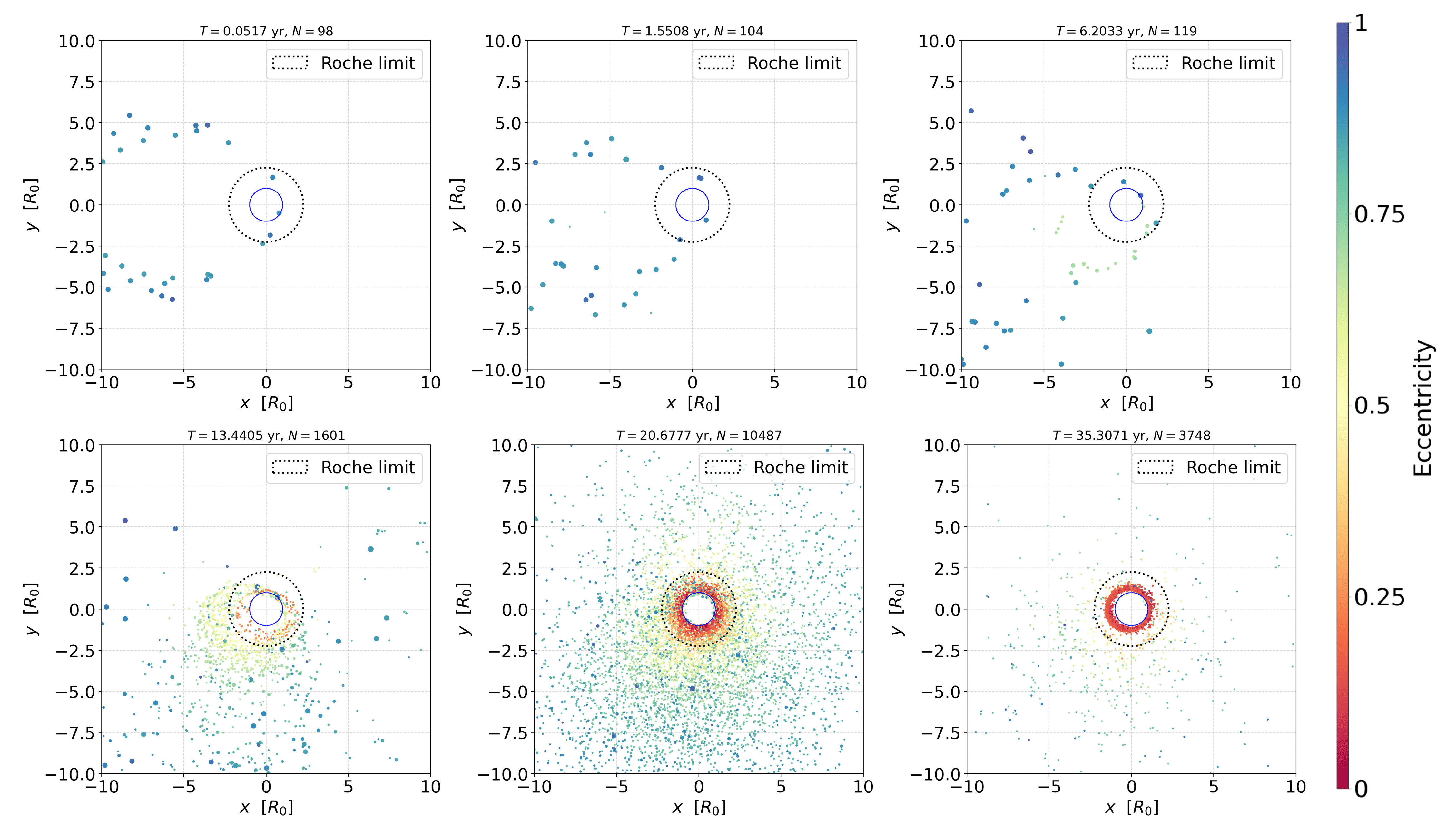}
\caption{The zoom-in view of Figure \ref{fig:snap_i45} (Model 4), but the color of each plot represents the particles eccentricity and their size reflect the particle radius.}
\label{fig:snap_i45_zoom}
\end{figure}

\begin{figure}
\centering
\includegraphics[width=\linewidth]{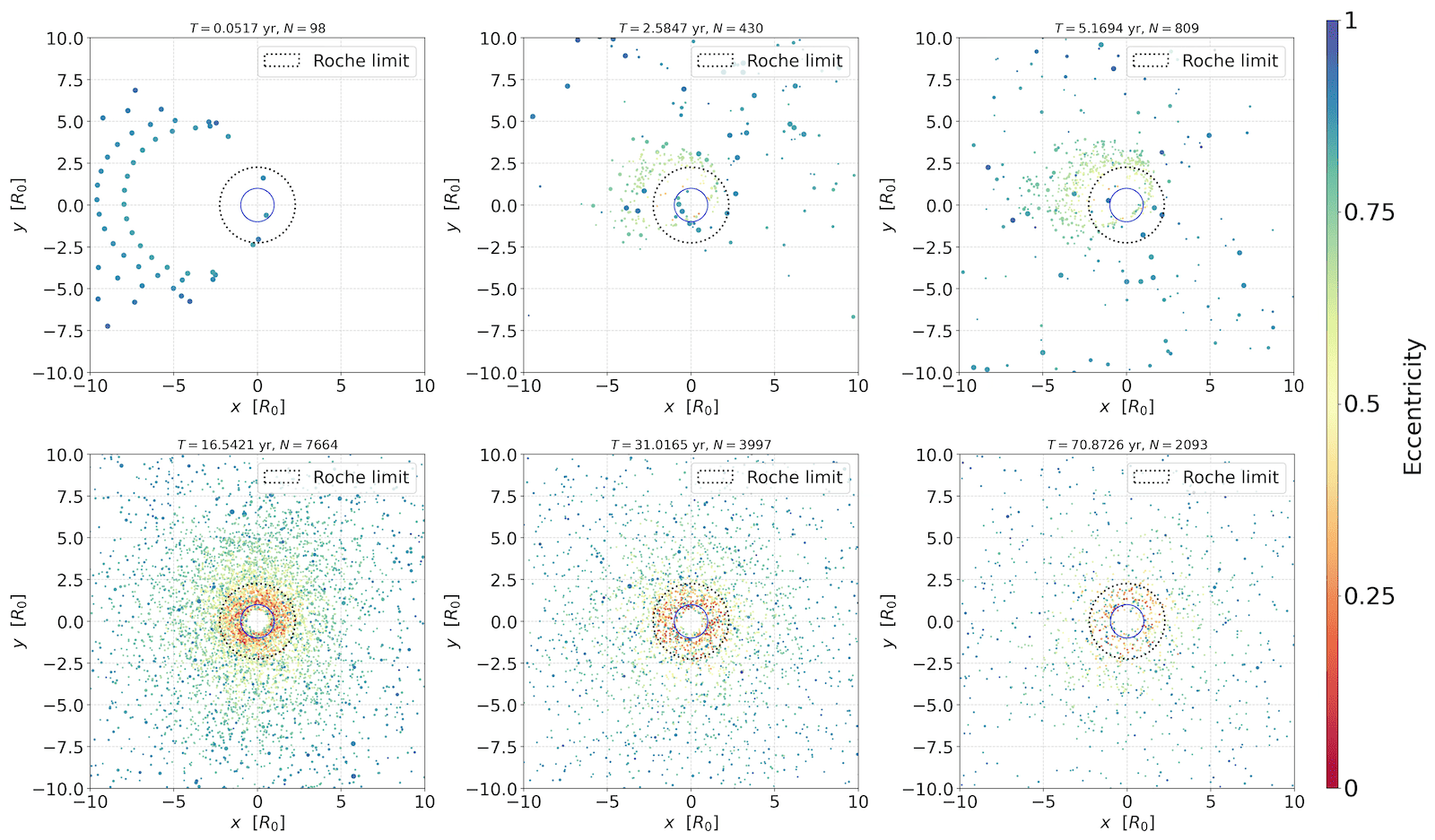}
\caption{The zoom-in view of Figure \ref{fig:snap_i60} (Model 5), but the color of each plot represents the particles eccentricity and their size reflect the particle radius.}
\label{fig:snap_i60_zoom}
\end{figure}

\begin{figure}
\centering
\includegraphics[width=\linewidth]{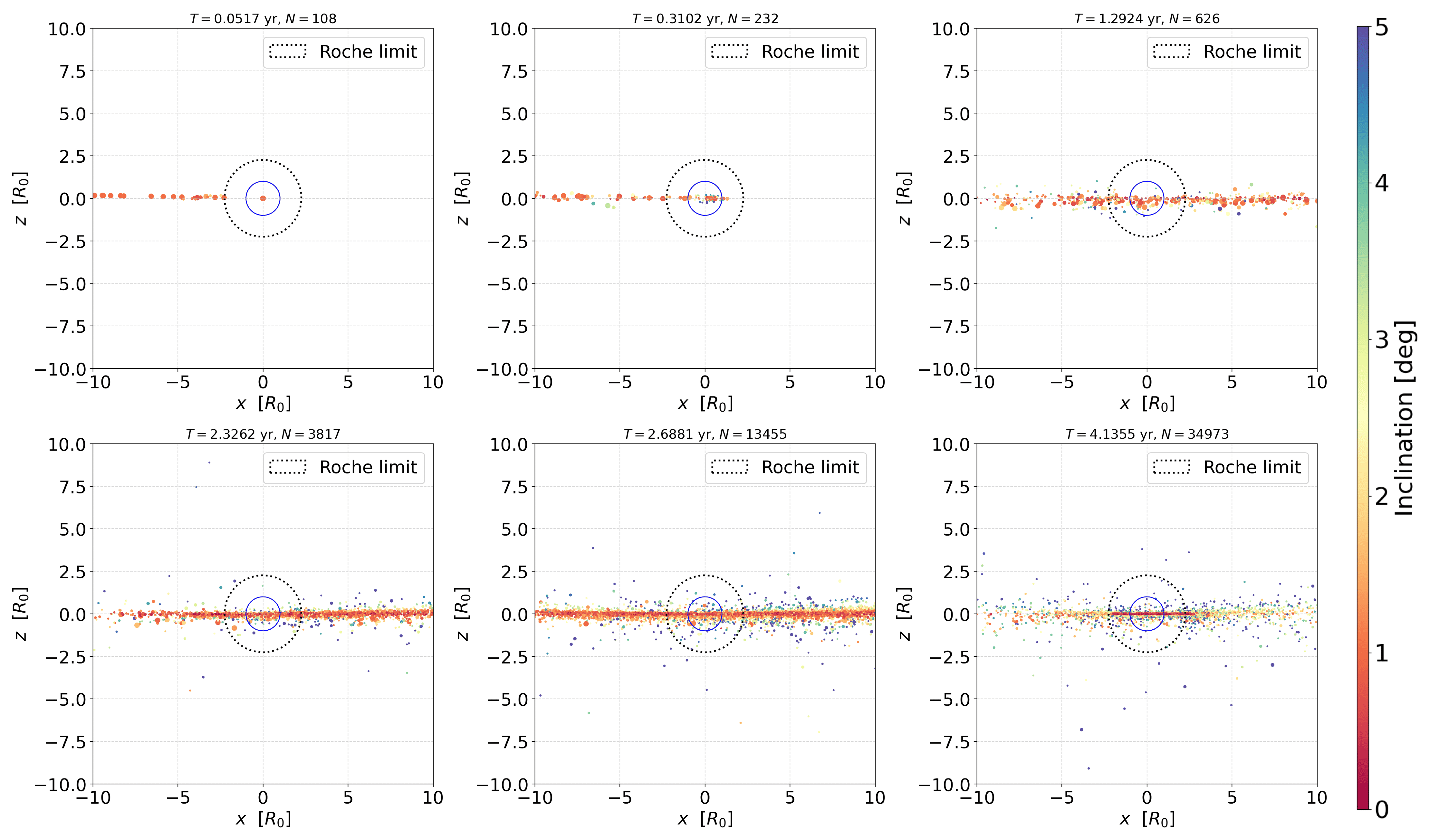}
\caption{The zoom-in side view of Figure \ref{fig:snap_i1.0} (Model 1), but the color of each plot represents the particles inclination and their size reflects the particle radius.}
\label{fig:snap_i1.0_zoom_z}
\end{figure}

\begin{figure}
\centering
\includegraphics[width=\linewidth]{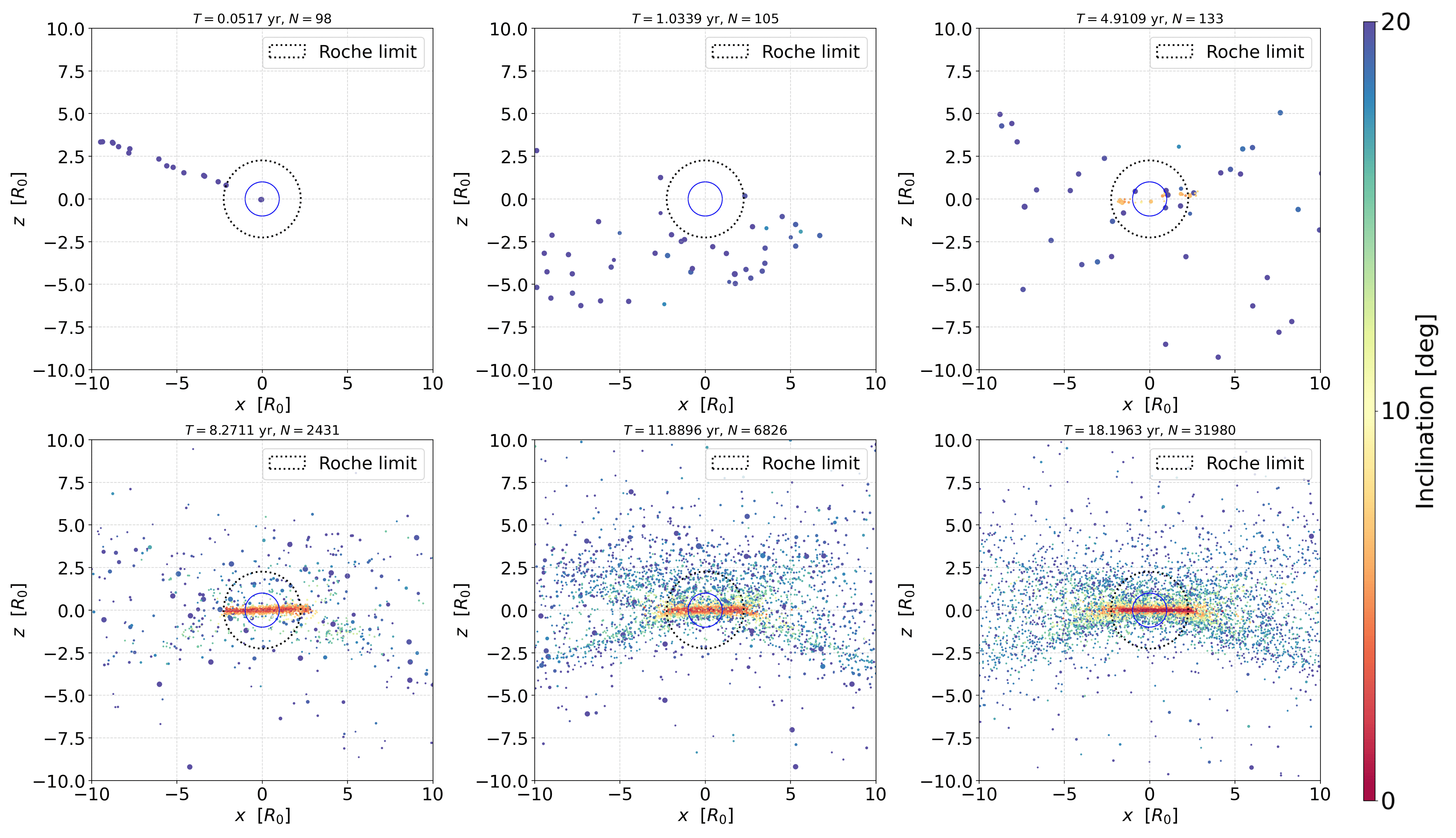}
\caption{The zoom-in side view of Figure \ref{fig:snap_i20} (Model 2), but the color of each plot represents the particles inclination and their size reflects the particle radius.}
\label{fig:snap_i20_zoom_z}
\end{figure}

\begin{figure}
\centering
\includegraphics[width=\linewidth]{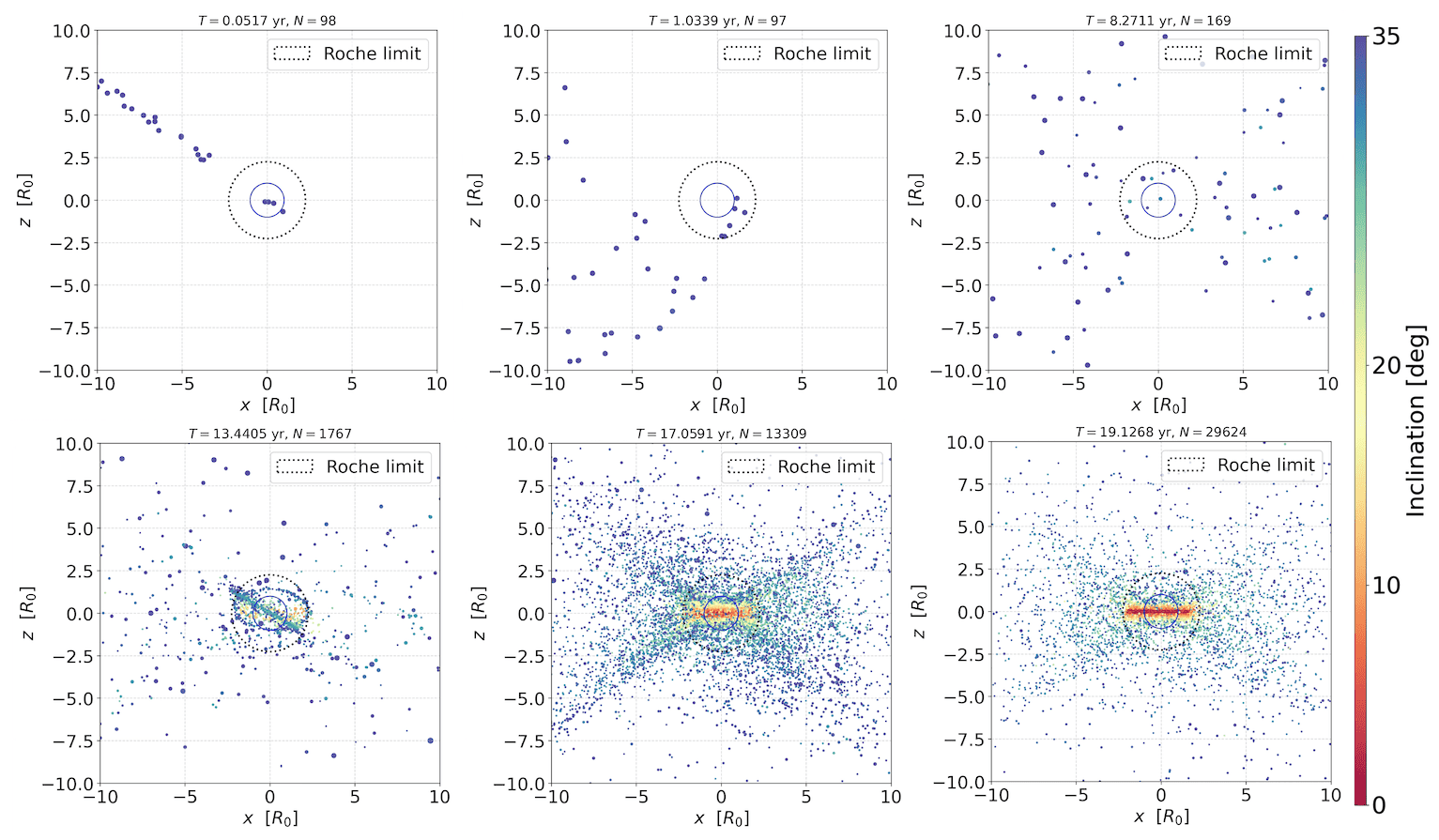}
\caption{The zoom-in side view of Figure \ref{fig:snap_i35} (Model 3), but the color of each plot represents the particles inclination and their size reflects the particle radius.}
\label{fig:snap_i35_zoom_z}
\end{figure}

\begin{figure}
\centering
\includegraphics[width=\linewidth]{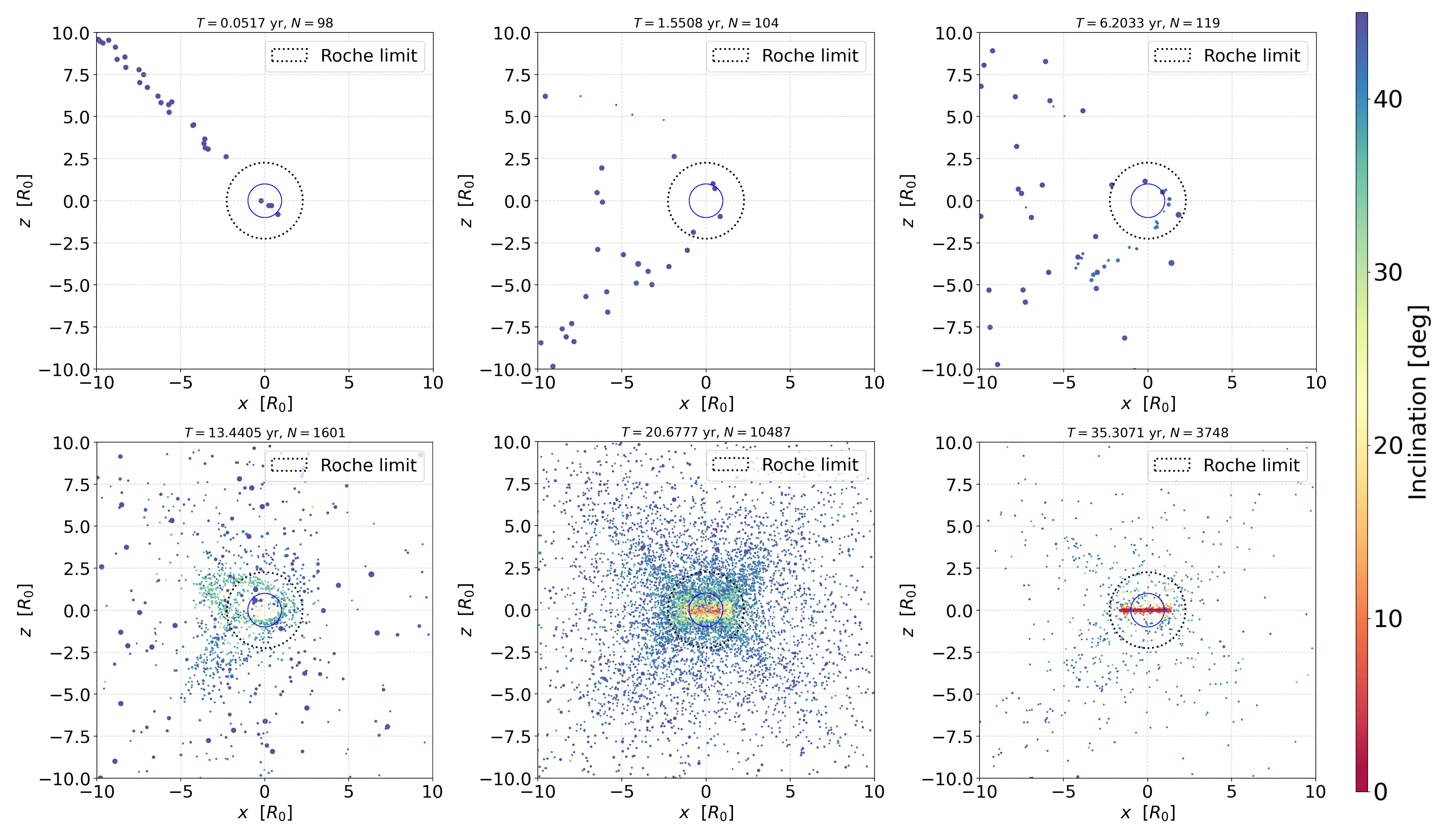}
\caption{The zoom-in side view of Figure \ref{fig:snap_i45} (Model 4), but the color of each plot represents the particles inclination and their size reflects the particle radius.}
\label{fig:snap_i45_zoom_z}
\end{figure}

\begin{figure}
\centering
\includegraphics[width=\linewidth]{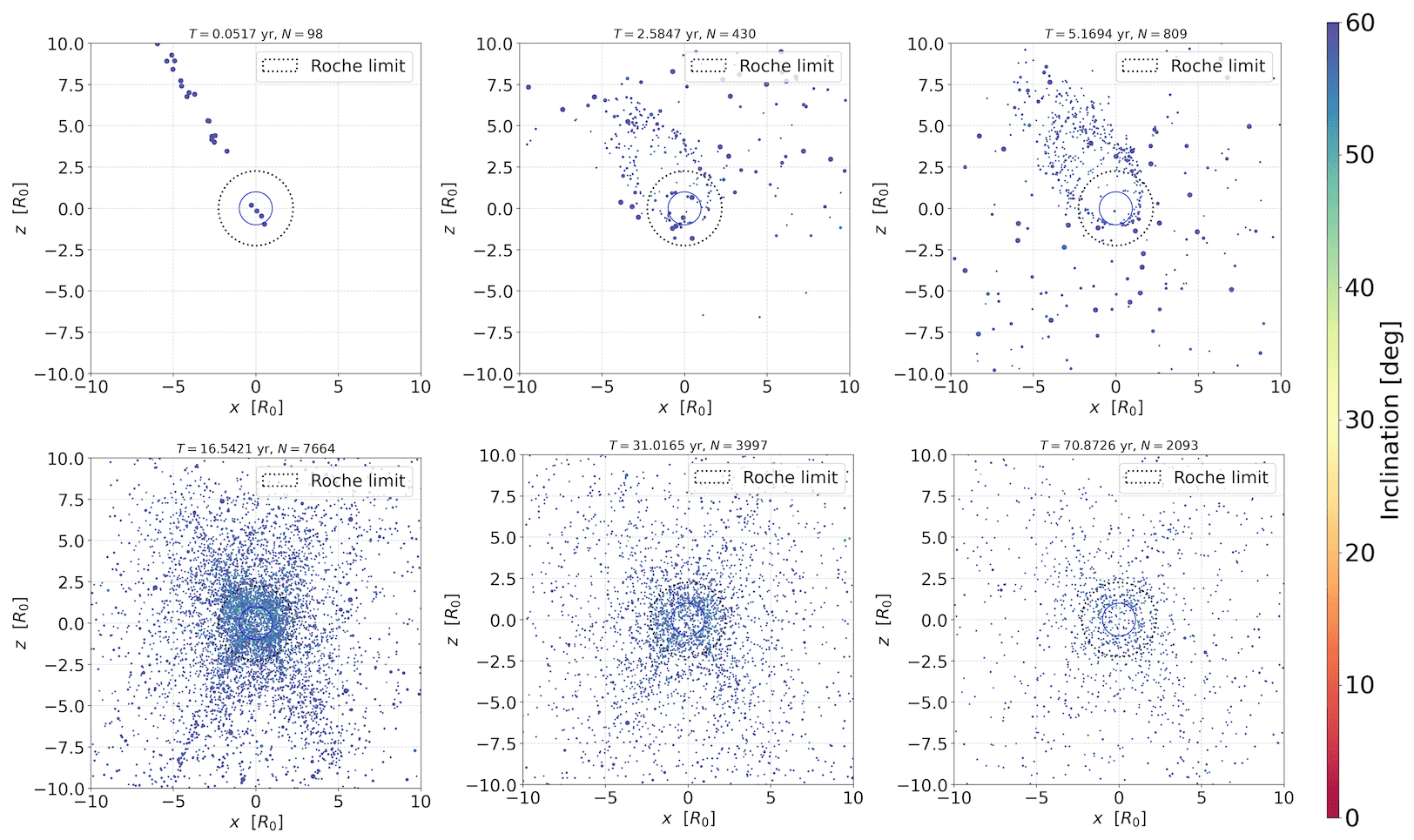}
\caption{The zoom-in side view of Figure \ref{fig:snap_i60} (Model 5), but the color of each plot represents the particles inclination and their size reflects the particle radius.}
\label{fig:snap_i60_zoom_z}
\end{figure}

Here, we present the additional snapshots.
Figures \ref{fig:snap_i1.0_zoom}, \ref{fig:snap_i20_zoom}, \ref{fig:snap_i45_zoom} and \ref{fig:snap_i60_zoom} show the zoom-in snap around Saturn in Model 1, 2, 4 and 5, respectively.
The color and size of each dots represent the particles eccentricity and their physical radius, respectively.
Figures \ref{fig:snap_i1.0_zoom_z}, \ref{fig:snap_i20_zoom_z}, \ref{fig:snap_i45_zoom_z} and \ref{fig:snap_i60_zoom_z} show the zoom-in snap from the side view around Saturn in Model 1, 2, 4 and 5, respectively.
The color and size of each dots represent the particles inclination and their physical radius, respectively.

\section*{Appendix B. The viscous timescale of the narrow ring}\label{app:timescale}
Here, we evaluate the viscous timescale to understand how fast the narrow rings which formed as a result of the collisional evolution diffuse by ring viscosity in the real system.
The ring viscosity of the self-gravitating rings can be calculated with \citep{Daisaka2001}
\begin{equation}
    \nu=26r_{\rm h}^{*5}\frac{G^2\Sigma^2}{\Omega^3},
\end{equation}
where $r_{\rm h}^*=r_{\rm h}/(2r_{\rm p})$, $\Sigma$ is the surface density, $\Omega=\sqrt{GM_0/r^3}$ is the orbital frequency, $r_{\rm p}$ is the particle radius and $r$ is the distance from the planet.
The particle's mutual Hill radius $r_{\rm h}$ is defined as 
\begin{equation}
    r_{\rm h}=\qty(\frac{2m_{\rm p}}{3M_0})^{1/3}r, \label{eq:rh}
\end{equation}
where $m_{\rm p}$ is the particle mass.
Thus, we can rewrite $r_{\rm h}^*$ from Eq. \ref{eq:Roche} and Eq. \ref{eq:rh}:
\begin{equation}
    r_{\rm h}^*=1.072\qty(\frac{r}{r_{\rm R}}).
\end{equation}
From these equations, we can evaluate the viscosity value of the narrow circular rings and its viscous diffusion timescale $\tau_{\rm vis}\sim(\Delta r)^2/\nu$, where $\Delta r$ is the radial width of the narrow ring.
We can write it as follow:
\begin{equation}
    \tau_{\rm vis}\simeq\frac{1}{3.72\Omega}\qty(\frac{r}{r_{\rm R}})^{-7}\qty(\frac{M_{\rm ring}}{M_0})^{-2}\qty[\frac{r_{\rm out}^2-r_{\rm in}^2}{r_{\rm R}^2}]^2\qty(\frac{\Delta r}{r_{\rm R}})^2,
\end{equation}
where $M_{\rm ring}$ is the mass of the narrow ring, $r_{\rm in}$ and $r_{\rm out}$ are its radius of inner and outer edge and $\Delta r=r_{\rm out}-r_{\rm in}$.
When we consider the narrow ring whose radius is $r=r_{\rm R}$, radial width is $\Delta r=0.2r_{\rm R}$ and mass is same as the captured mass $M_{\rm ring}=10^{22} \ {\rm kg}\sim1.75\times10^{-5}M_0$, we can estimate the viscous timescale is $\tau_{\rm vis}\sim 1400$ years.
On the other hand, the damping timescale of the eccentricity and inclination leading to the formation of the circular equatorial rings $\tau_{\rm ring}$ is order of $\sim 10$ years (see Fig. \ref{fig:snap_i1.0}, \ref{fig:snap_i20} and \ref{fig:snap_i35}) corresponding to $\sim 660 T_{\rm K}$, where $T_{\rm K}$ is Kepler orbital period at the semimajor axis $a=10 R_0$.
Thus, the dynamical timescale of eccentricity and inclination damping is much smaller than viscous timescale $\tau_{\rm ring}\ll\tau_{\rm vis}$.

\end{document}